\documentclass[preprint,12pt,letterpaper]{elsarticle}

\usepackage{amssymb}
\usepackage{amsthm}
\usepackage{amsmath, mathrsfs}
\usepackage{algorithm}
\usepackage{algorithmicx}
\usepackage{algpseudocode}
\usepackage{color}
\usepackage{ulem} 

\biboptions{sort&compress}
\newdefinition{rmk}{Remark}

\journal{Journal of Computational Physics}

\newtheorem{dfn}{Definition}
\newcommand{\g}{\ensuremath{_\Gamma}}
\newcommand{\gn}[1]{\ensuremath{_{\Gamma_{#1}}}}
\newcommand{\ord}[1]{\ensuremath{O(#1)}}

\newcommand{\x}{\ensuremath{\vec{x}}}
\newcommand{\xq}{\ensuremath{\vec{x}_q}}
\newcommand{\xg}{\ensuremath{\mathcal{X}_{\Gamma}}}

\newcommand{\sq}{\ensuremath{\mathcal{S}(\xq, \ell_q)}}
\newcommand{\cb}{\ensuremath{\mathcal{C}(\xq, \ell_q)}}
\newcommand{\myremarkend}{~\hfill$\clubsuit\/$}

\DeclareMathOperator{\argmin}{arg\, min}

\begin{document}

\begin{frontmatter}


\title{Imposing jump conditions on nonconforming interfaces for the Correction Function Method: a least squares approach}


\author[mit-aa]{Alexandre Noll Marques}
\author[mcgill]{Jean-Christophe Nave}
\author[mit-math]{Rodolfo Ruben Rosales}

\address[mit-aa]{Department of Aeronautics and Astronautics,
                 Massachusetts Institute of Technology\\
                 Cambridge, MA 02139-4307}
\address[mcgill]{Department of Mathematics and Statistics,
                 McGill University\\
                 Montreal, Quebec H3A 0B9, Canada}
\address[mit-math]{Department of Mathematics,
                   Massachusetts Institute of Technology\\
                   Cambridge, MA 02139-4307}

\begin{abstract}
We introduce a technique that simplifies the problem of
imposing jump conditions on interfaces that are not aligned
with a computational grid in the context of the
\textit{Correction Function Method} (CFM).
The CFM offers a general framework to solve Poisson's
equation in the presence of discontinuities to high order
of accuracy, while using a compact discretization stencil.
A key concept behind the CFM is enforcing the jump
conditions in a least squares sense.
This concept requires computing integrals over sections of
the interface, which is a challenge in 3-D when only an
implicit representation of the interface is available (e.g.,
the zero contour of a level set function).
The technique introduced here is based on a new formulation
of the least squares procedure that relies only on integrals
over domains that are amenable to simple quadrature after
local coordinate transformations.
We incorporate this technique into a fourth order accurate
implementation of the CFM, and show examples of solutions to
Poisson's equation computed in 2-D and 3-D.
\end{abstract}


\end{frontmatter}



\section{Introduction} \label{sec:intro}
%
Solving Poisson's equation in the presence of
discontinuities is of great importance in science and
engineering applications.
In many cases, the discontinuities are caused by interfaces
between different media, such as in multiphase flows,
Stefan's problem, Janus drops, and other multiphase
phenomena.
These interfaces follow themselves from solutions to
differential equations, and can assume complex
configurations.
For this reason, it is convenient to embed the interface
into a regular triangulation or Cartesian grid and solve
Poisson's equation in this regular domain.
The Correction Function Method (CFM)~\citep{marques:2011,
marques:2012, marques:2017} was developed to solve Poisson's
equation in this context, and achieve high order of
accuracy with a compact discretization stencil.

The CFM is based on the concept of the correction function
(introduced in \S\ref{sub:cf}), which is the minimizer of an
energy functional constructed to impose the jump conditions
and locally satisfy Poisson's equation.
In the present paper we leverage the flexibility in this
construction and propose a new formulation that simplifies
numerical implementation.
A key ingredient of the CFM is computing integrals over
sections of the interface, which is a challenge in 3-D when
only an implicit representation of the interface is
available (e.g., the zero contour of a level set function).
There are several recent methods that focus specifically on
integrating functions on implicitly defined
surfaces~\cite{wen:2010, muller:2013, saye:2015,
schwartz:2015, fries:2016}.
However, these methods are designed to solve general
problems that include non-trivial integration domains.
In our context, we are free to define the energy functional
and we use this flexibility to simplify the integration
problem.
Specifically, we define integration domains using
approximate coordinate transformations such that we can
always use standard quadrature techniques.

\begin{figure}[htb!]
 \begin{center}
  \includegraphics[width=2in]{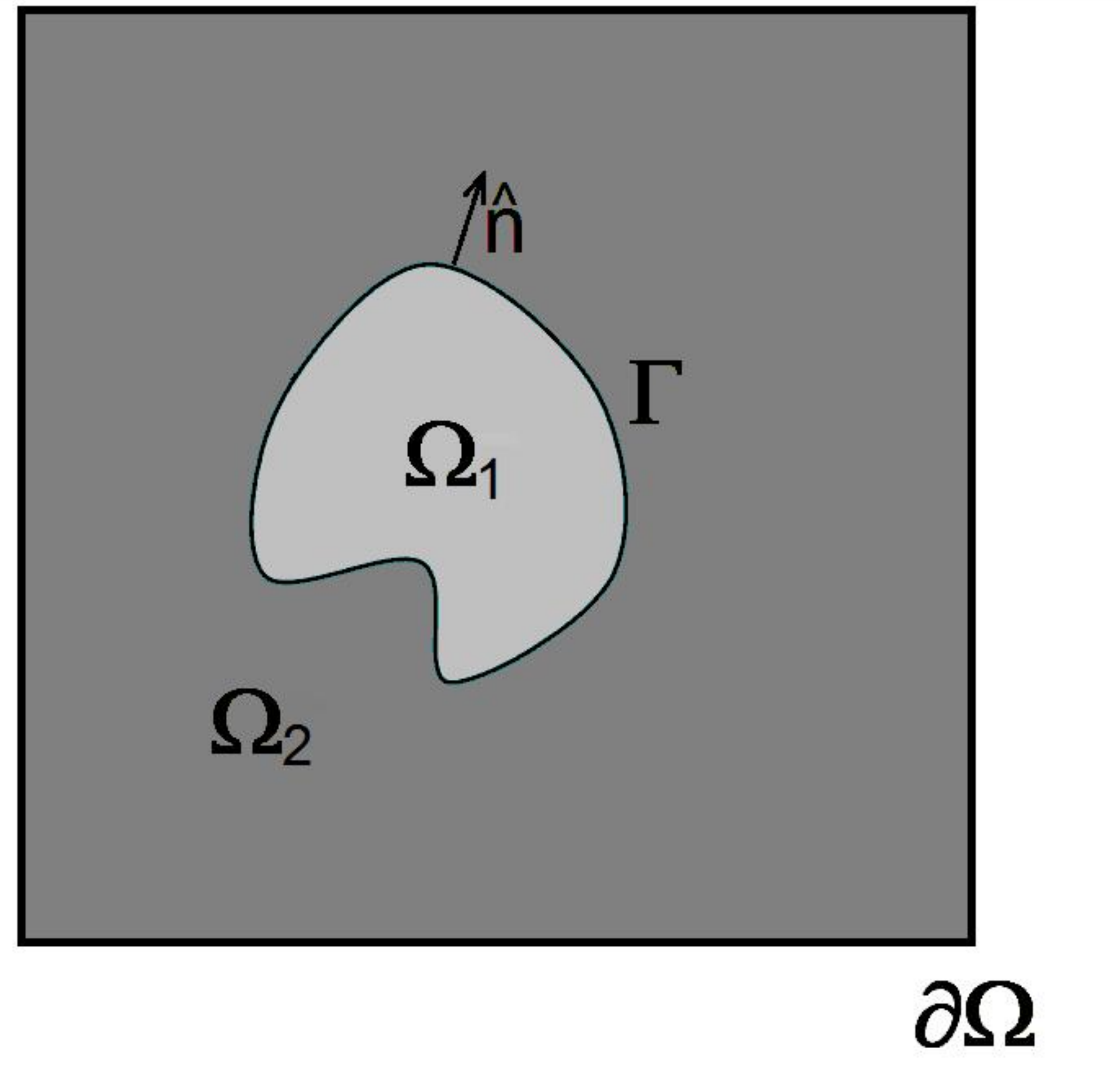}
 \end{center}
 \caption{Illustration of the solution domain for the
          discontinuous Poisson equation.
          The domain $\Omega\/$ is split into two
          subdomains, $\Omega_1\/$ and $\Omega_2\/$, by the
          internal interface $\Gamma\/$.}
 \label{fig:problem}
\end{figure}

We now define the problem precisely.
Poisson's equation with imposed discontinuities (for
brevity: discontinuous Poisson equation) is given by
%
\begin{subequations}\label{eq:poisson}
 \begin{align}
   \nabla \cdot \bigl(\beta(\vec{x}) \nabla u(\vec{x})\bigr)
   &= f(\vec{x})
   &\mathrm{for}\;\; \vec{x}
   &\in \Omega\/,
   \label{eq:poisson-eq}\\
   [u(\vec{x})] 
   &= a(\vec{x})
   &\mathrm{for}\;\; \vec{x}
   &\in \Gamma\/,
   \label{eq:a}\\
   [\beta(\vec{x}) u_n(\vec{x})]
   &= b(\vec{x})
   &\mathrm{for}\;\; \vec{x}
   &\in \Gamma\/,
   \label{eq:b}\\
   u(\vec{x})
   &= g(\vec{x})
   &\mathrm{for}\;\; \vec{x}
   &\in \partial\Omega\/.
   \label{eq:dirichlet}
 \end{align}
\end{subequations}
%
Here the solution domain $\Omega\/$ is split into two
sub-domains, $\Omega_1\/$ and $\Omega_2\/$, by a
co-dimension 1 surface, $\Gamma\/$, disjoint from the
boundary $\partial \Omega\/$.
This situation is illustrated in
figure~\ref{fig:problem}.
Furthermore, $a\/$ and $b\/$ are known functions defined on
$\Gamma\/$, and square brackets denote the jump in the
enclosed quantity across $\Gamma\/$, i.e.,
%
\begin{equation*}
 [u(\vec{x})]
 = \lim_{\substack{  \vec{x}^{\ast} \, \to \, \vec{x}
                  \\ \vec{x}^{\ast} \, \in \, \Omega_2}}
   u(\vec{x}^{\ast})
 - \lim_{\substack{  \vec{x}^{\ast} \, \to \, \vec{x}
                  \\ \vec{x}^{\ast} \, \in \, \Omega_1}}
   u(\vec{x}^{\ast})\/.
\end{equation*}

In addition, we assume Dirichlet conditions on the domain
boundary $\partial \Omega\/$, and that this boundary is
aligned with the computational grid.
However, other boundary conditions can be used, such as
Neumann or Robin, with no effect on how jump conditions are
imposed on the interface $\Gamma\/$.

\begin{rmk}\label{rmk:beta}
The coefficient $\beta\/$ denotes a known positive function,
$\beta = \beta(\x)$.
In many applications, $\beta\/$ is discontinuous across the
interface $\Gamma\/$.
However, in this paper we only consider the case where
$\beta\/$ is constant.
The subject of this paper is imposing the jump conditions when
only an implicit representation of the interface is
available.
The case of discontinuous $\beta\/$ introduces additional
difficulties in the context of the CFM formulation that are
not related to the subject of this paper.
Hence, we address the matter of discontinuous $\beta\/$ in
separate work.
This matter was partially addressed in
ref.~\citep{marques:2017} by coupling the CFM with boundary
integral equations.
However, solving boundary integral equations to high order
of accuracy using level set functions is also a challenge.
For this reason, work on a more general framework for the
case of discontinuous $\beta\/$ is ongoing and will be
addressed in future publications.
\end{rmk}

Over the past four decades, several methods have been
developed to solve~\eqref{eq:poisson}, and other closely
related problems, with either an interface or a boundary
that is embedded into a regular triangulation or Cartesian
%
%
%
grid~\citep{peskin:1972, peskin:1977, peskin:1993,
goldstein:1993, lai:2000, cortez:2000, peskin:2002,
griffith:2005, mittal:2005,
%
%
stein:2016,
%
%
johansen:1998,
%
%
mayo:1984, mayo:1992, mckenney:1995,
%
%
leveque:1994, leveque:1997, li:2001, lee:2003, linnick:2005,
zhong:2007,
%
%
fedkiw:1999, liu:2000, kang:2000,
%
%
gibou:2005, gibou:2007,
%
zhou:2006,
%
%
%
guittet:2015,
%
li:2003, hou:2005, hou:2013}.
One of the shortcomings of many embedded methods is that
they are rather hard to generalize beyond first or second
order of accuracy.
However, there are methods that achieve accuracy higher than
second order, based on one of the following ideas:
(i)~discretization stencils that incorporate the jump (or
    boundary) conditions~\citep{mayo:1984, mayo:1992,
    mckenney:1995, linnick:2005, zhong:2007}, or
(ii)~smooth extrapolations of the solution constrained by
     the jump (or boundary) conditions~\citep{stein:2016,
     gibou:2005, gibou:2007, zhou:2006}.
In practice these ideas are implemented by Taylor expansion
or a similar concept, and high order of accuracy is obtained
at the expense of wide discretization stencils.
In turn, wide stencils introduce additional issues, such as
handling multiple crossings of the interface by a single
stencil, and restrictions on the proximity between
interfaces.
The method introduced by Mayo and
collaborators~\citep{mayo:1984, mayo:1992, mckenney:1995}
avoids wide discretization stencils by incorporating second
and third derivatives of the jump conditions into the Taylor
expansion.
On the other hand, computing higher derivatives of the jump
conditions requires the solution of an additional boundary
integral equation.

In contrast to using Taylor expansion, the
CFM~\citep{marques:2011, marques:2012, marques:2017} is
based on computing a smooth extension of the solution by
solving a partial differential equation that is compatible
with Poisson's equation.
This concept results in a general framework that, in
principle, can achieve arbitrary order of accuracy,
and maintains compact discretization stencils.
In ref.~\citep{marques:2011} we introduced the fundamentals
of the CFM, and proposed a fourth order implementation to
solve \eqref{eq:poisson} in 2-D when $\beta\/$ is constant.
In ref.~\citep{marques:2017} we extended the CFM to
solve~\eqref{eq:poisson} with piece-wise constant $\beta\/$,
including the possibility of arbitrarily large jumps in the
equation's coefficients
(in~\citep{marques:2017} we showed third order convergence
for coefficient ratios of \ord{10^6}).
The work of ref.~\citep{marques:2017} is inspired by
Mayo's method~\citep{mayo:1984, mayo:1992}, and also
requires the solution of a boundary integral equation.
However, there is an extension of the CFM for the case of
discontinuous $\beta\/$ that does not require the solution
of a boundary integral equation.
This extension is briefly discussed in
ref.~\citep{marques:2012}, and is the subject of current
research by the authors.
The CFM has also been extended to other classes of
differential equations, such as the heat
equation~\citep{marques:2012}, the Navier-Stokes
equations~\citep{marques:2012}, and the wave
equation~\citep{abraham:2018}.

This paper is organized as follows.
In \S\ref{sec:cfm} we present an overview of the CFM for
Poisson's equation.
Next, in \S\ref{sec:integrals} we present the details of the
new technique for imposing the jump conditions in a least
squares sense, including its effects on the CFM formulation.
In \S\ref{sec:results} we present solutions computed with
the modified CFM in 2-D and 3-D.
Finally, the conclusions are in \S\ref{sec:conclusion}.

\section{Overview of the Correction Function Method}
\label{sec:cfm}
%

\subsection{The correction function}
\label{sub:cf}
The Correction Function Method (CFM) was developed to solve
the discontinuous Poisson problem \eqref{eq:poisson} when
the interface $\Gamma\/$ is not aligned with a computational
grid~\citep{marques:2011, marques:2012, marques:2017}.
The CFM is based on the notion of smooth extensions of the
solution (denoted by $u_1\/$ and $u_2\/$) across the
interface, and the definition of the
\textit{correction function}: $D = u_2 - u_1\/$.
This function can be used to complete discretization
stencils that stride the interface.

\begin{figure}[b!] 
 \begin{center}
  \includegraphics[width=3in]{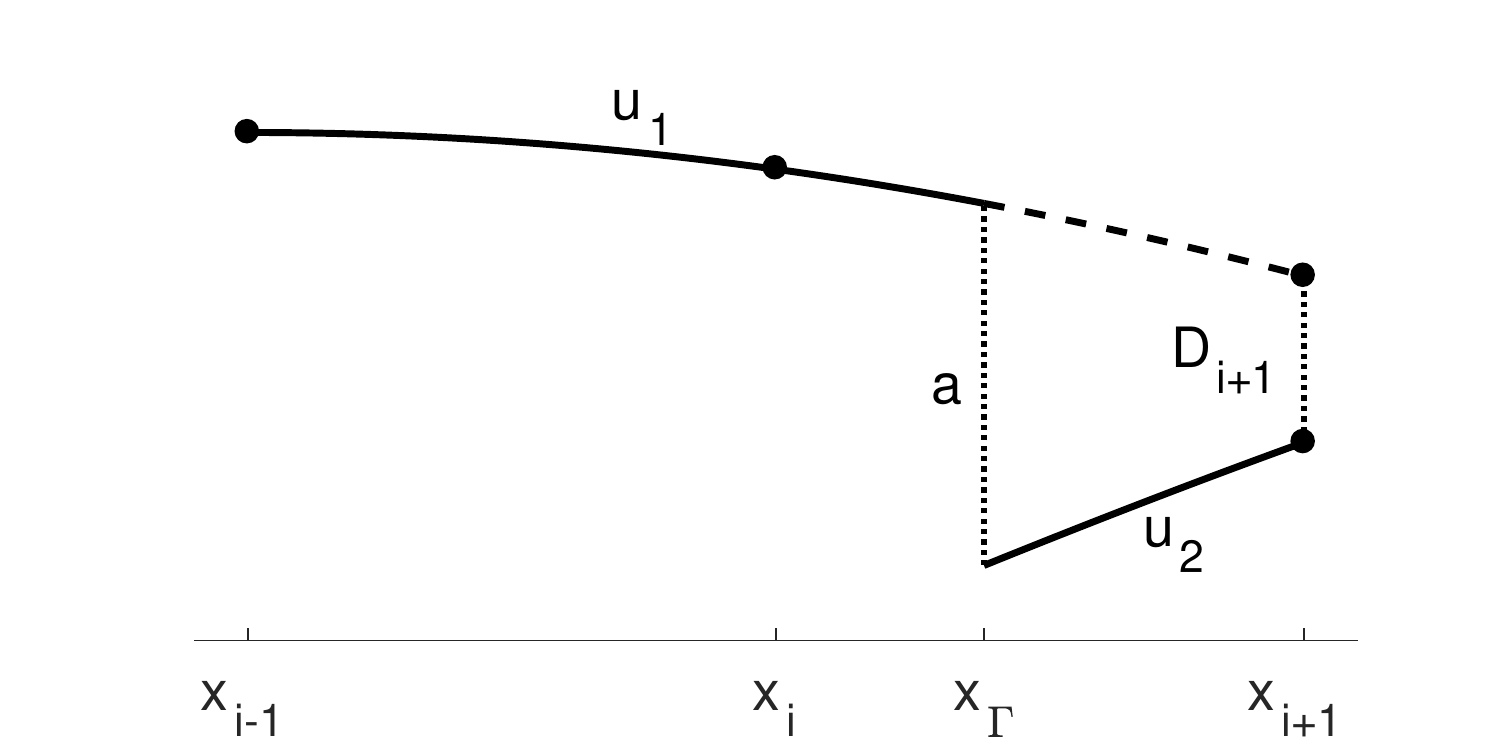}
 \end{center}
 \caption{Illustration of how the correction function is
          used to complete discretization stencils. To
          compute $u_{xx_i}\/$, write
          $u_{1_{i+1}} = u_{2_{i+1}} - D_{i+1}\/$, which
          leads to~\eqref{eq:uxx_D}.}
 \label{fig:uxx}
\end{figure}

We illustrate this point with a 1-D example.
Consider the approximate computation of $u_{xx}\/$ at grid
node $i\/$ using standard centered finite differences:
\begin{equation}\label{eq:uxx}
 u_{xx_i} \approx \dfrac{u_{i-1} - 2u_i + u_{i+1}}{h^2},
\end{equation}
where $h\/$ is a uniform grid spacing.
Equation~\eqref{eq:uxx} is known to produce errors
$O(h^2)\/$ if $u \in C^2((x_{i-1},x_{i+1}))\/$.
However, when $u\/$ is discontinuous, as depicted in
figure~\ref{fig:uxx}, the approximation in \eqref{eq:uxx} is
not valid since it is based on Taylor expansions.
One way to address this issue, originally introduced in the
Ghost Fluid Method~\citep{fedkiw:1999, liu:2000, kang:2000},
is to estimate a smooth extension of the solution across the
interface before applying the discretization.
In practice, only the difference between the smooth
extension and the actual grid values are needed.
In the case depicted in figure~\ref{fig:uxx}, an estimate of
$D_{i+1} = u_{2_{i+1}} - u_{1_{i+1}}\/$ can be used to
correct~\eqref{eq:uxx} as follows:
\begin{equation}\label{eq:uxx_D}
 u_{xx_i} \approx \dfrac{u_{i-1} - 2u_i + u_{i+1}}{h^2} 
                - \dfrac{D_{i+1}}{h^2}.
\end{equation}
To achieve high order of accuracy, the correction $D\/$
needs to be computed to at least the same order of accuracy
as the discretization of Poisson's equation.
Other methods use Taylor expansions to estimate
$D\/$~\citep{liu:2000, kang:2000} or, equivalently, smooth
extrapolations of $u\/$~\citep{fedkiw:1999, gibou:2005,
gibou:2007, zhou:2006}.
In contrast, the CFM~\cite{marques:2011} computes $D\/$ by
solving a partial differential equation defined on a narrow
band of width $O(h)\/$ that surrounds the interface.
As shown in ref.~\citep{marques:2011}, when $\beta\/$ is
constant, the correction function is defined as the solution
to the following equation,
%
\begin{subequations}\label{eq:D}
 \begin{align}
  \nabla^2 D(\x)
  &= \bigl(f_2(\x) - f_1(\x)\bigr)/\beta
   = f_D(\x)
  &\mathrm{for}\;\; \x
  &\in \Omega_{\Gamma},
  \label{eq:D-poisson}\\
  D(\x)          &= a(\x)
                 &\mathrm{for}\;\; \x
                 &\in \Gamma,
  \label{eq:D-a}\\
  D_n(\x)        &= b(\x)
                 &\mathrm{for}\;\; \x
                 &\in \Gamma\/,
 \label{eq:D-b}
 \end{align}
\end{subequations}
where $\Omega_{\Gamma}\/$ denotes the narrow band around the
interface in which the correction function is defined.

In \eqref{eq:D}, we assume that we have access to smooth
extensions of $f_1\/$ and $f_2\/$ across the interface.
To maintain the accuracy of the computations, these
extensions must in $C^{m-2}\/$, where $m\/$ denotes the
desired order of accuracy.
In most practical applications, $f_1\/$ and $f_2\/$ are
simple functions (e.g., a constant), and computing the
extensions is trivial.
In more complex cases, the extensions can be computed
following, for instance, the algorithm described by
Aslam~\citep{aslam:2004}.

\begin{rmk}\label{rmk:rhs}
 Because~\eqref{eq:D} depends only on known parameters of
 the problem ($a\/$, $b\/$, $f\/$, and the position of
 $\Gamma$), the CFM produces modifications to the
 right-hand-side of the discretized equations only.
 As a result, one can solve Poisson's equation by
 inverting the exact same linear system as in problems with
 no interface, but with a modified right-hand-side.
\end{rmk}

\begin{rmk}\label{rmk:cauchy}
 Equation~\eqref{eq:D} is an elliptic Cauchy problem.
 In a continuous setting, this problem is ill-posed because
 small perturbations to the interface conditions
 (\ref{eq:D-a}-c) can result in arbitrarily large changes in
 the solution.
 However, in a numerical setting, where disturbances to 
 (\ref{eq:D-a}-c) are restricted to a finite wave length, it
 is possible to develop well-behaved numerical schemes to
 solve this problem in a narrow band surrounding the
 interface.
 The least squares approach used in the CFM (discussed
 below) is one such numerical scheme.
 In ref.~\citep{marques:2011} we explain the implications of
 using this method in terms of conditioning.\myremarkend
\end{rmk}

In practice, it is convenient to solve \eqref{eq:D} locally
whenever the stencil used to discretize the Laplace operator
in~\eqref{eq:poisson-eq} crosses the interface.
Namely, \eqref{eq:D} is solved in small patches
$\Omega_{\Gamma_i}\/$, $i \in S_{\Gamma}\/$, where
$S_{\Gamma}\/$ denotes the set of stencils that cross the
interface.
The construction of these patches is described in
\S\ref{sub:patch}.

The correction function is computed by solving a least
squares minimization.
Let $P_m(\Omega_{\Gamma_i})\/$ denote some class of
approximating functions that allow $m^{th}$ order
approximations to smooth functions within
$\Omega_{\Gamma_i}\/$.
For example, polynomials of order $m-1\/$.
We search for an approximate solution to~\eqref{eq:D},
within $\Omega_{\Gamma_i}\/$, of the form
$\tilde{D}_i(\x; \, \vec{c}) = \sum_j c_j \varphi_j(\x)\/$,
$\varphi_j \in P_m(\Omega_{\Gamma_i})\/$.
The coefficients $c_j\/$ are computed by solving the
following minimization problem.
%
\begin{equation}
 \label{eq:ls_minimization}
 \vec{c}^{\ast} = \argmin \, J_i(\vec{c})\,
\end{equation}
where
\begin{equation}
 \label{eq:J}
 \begin{split}
  J_i (\vec{c}) &=
  \dfrac{\ell_i^4}{V_i}
  \int_{\Omega_{\Gamma_i}}
   \bigl( \nabla^2 \tilde{D}(\x; \, \vec{c}) - f_D(\x)
   \bigr)^2\, dV \\
  &+ \dfrac{1}{S_i}
  \int_{\Gamma_i}
  \Bigl( \bigl( \tilde{D}(\x; \, \vec{c}) - a(\x) \bigr)^2 +
         \ell_i^2 \bigl( \tilde{D}_n(\x; \, \vec{c}) - b(\x)
                  \bigr)^2
  \Bigr)\, dS\/.
 \end{split}
\end{equation}
%
In the expression above, $\ell_i$ denotes a characteristic
length of the patch $\Omega_{\Gamma_i}\/$, $V_i\/$ is the
volume of $\Omega_{\Gamma_i}\/$, $\Gamma_i\/$ is the
intersection of the interface $\Gamma\/$ with the patch
$\Omega_{\Gamma_i}\/$, and $S_i\/$ is the area of this
intersection.

The surface integral in~\eqref{eq:J} is responsible for
imposing the jump conditions from the original Poisson
problem.\footnote{%
Technically the surface integral in~\eqref{eq:J} imposes the
Cauchy boundary conditions~(\ref{eq:D}b--c).
But since these conditions are derived from the jump
conditions of the original Poisson problem, we refer to
(\ref{eq:D}b--c) as ``jump conditions'' throughout the
text.}
Nevertheless, evaluating these integrals is challenging in
3-D when only an implicit representation of the interface is
available.
The source of the problem is that correction function
patches can intersect the interface in an arbitrary fashion,
resulting in surface integrals that cannot be evaluated with
standard numerical quadrature.
In \S\ref{sec:integrals} we introduce a modification to the
least squares formulation that enables a relatively simple
implementation of the CFM in 3-D.

%
\subsection{Correction function patch}
\label{sub:patch}
%

\begin{figure}[t!] 
 \begin{center}
  \includegraphics[scale=0.3]{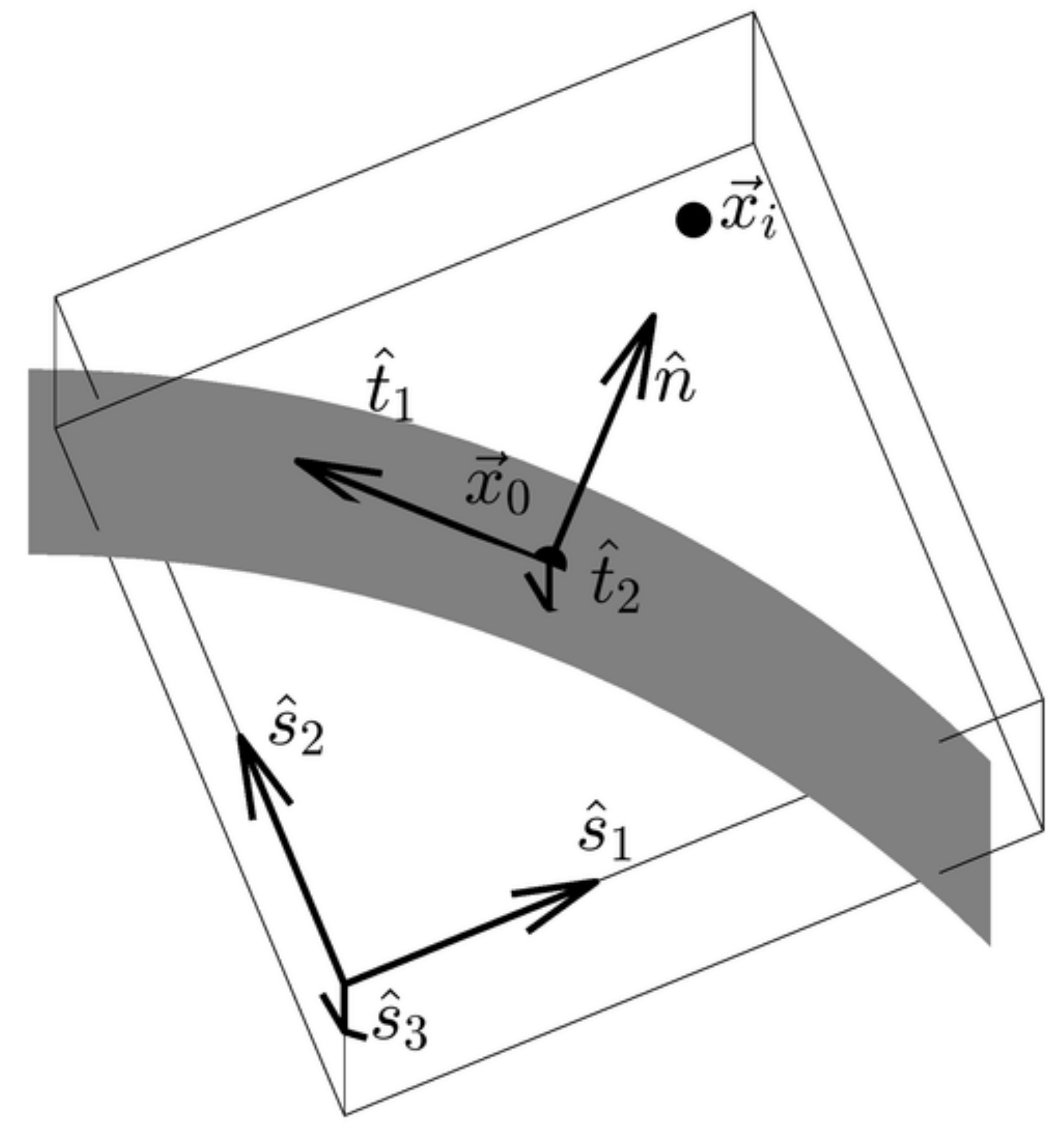}
 \end{center}
 \vspace{-0.1in}
 \caption{Patch used to define the
          \textit{correction function}.}
 \label{fig:patch3D}
\end{figure}

Here we use the ``node-centered'' \citep{marques:2011}
approach to construct the patches used to compute the
correction function, as illustrated in
figure~\ref{fig:patch3D} and described below.

Let us consider stencil $i \in S_{\Gamma}\/$, where
$S_{\Gamma}\/$ is the set of stencils that straddle the
interface.
We denote $\ell_i\/$ as the maximum distance between grid
points that are part of stencil $i$ (the characteristic
length of the stencil), and $\x_i\/$ as the average position
of those same grid points (the center position of the
stencil).
Furthermore, $P_{\Gamma}(\x)\/$ denotes an approximate
projection of \x\/ onto the interface $\Gamma\/$, defined
as
\begin{equation}\label{eq:projection}
 P_{\Gamma}(\x)
 = \x
 - \phi(\x)
   \Biggl(
    \dfrac{\vec{\nabla}\phi(\x)}
     {||\vec{\nabla}\phi(\x)||^2}\Biggr)\/,
\end{equation}
where $\phi\/$ is the level set function used to represent
the interface.

The correction function patch associated with stencil $i$,
denoted as $\Omega_{\Gamma_i}\/$, is a cube of edge length
$\ell_i$ centered at $\x_{0_i} = P_{\Gamma}(\x_i)\/$.
Furthermore, the patch is oriented such that one of its
diagonal planes coincides with the plane tangent to the
interface at $\x_{0_i}\/$.
Let
$\hat{n} = \vec{\nabla} \phi(\x_{0_i}) /
||\vec{\nabla} \phi(\x_{0_i})||\/$ denote the unit vector
normal to the interface at $\x_{0_i}\/$, and $\hat{t}_1\/$
and $\hat{t}_2\/$ denote vectors tangent to the surface,
given by
\begin{equation}\label{eq:tangent}
 \hat{t}_1 = \hat{n} \times \hat{e}_{\min},
 \quad \hat{t}_2 = \hat{n} \times \hat{t}_1\/,
\end{equation}
where $\hat{e}_{\min}\/$ corresponds to the coordinate
axis for which $(\hat{n} \cdot \hat{e}_{\min})\/$ is
minimum.
Then, define the following triad of unit vectors,
\begin{equation*}
 \hat{s}_1 = (\hat{n} - \hat{t}_1)/\sqrt{2}\/, \quad
 \hat{s}_2 = (\hat{t}_2 - \hat{s}_1)\/, \quad
 \hat{s}_3 = \hat{t}_2\/.
\end{equation*}
Finally, $\Omega_{\Gamma_i}\/$ is given by
\begin{equation*}
 \Omega_{\Gamma_i} =
  \left\{ \x \; \bigg \vert \;
     \x = \x_{0_i}
        + \dfrac{\ell_i}{2} \sum_{i=1}^3 \gamma_i \hat{s}_i, 
     \, \gamma_i \in [-1, 1] \right\}.
\end{equation*}

\subsection{Accuracy and robustness}
\label{sub:accuracy}

The accuracy of the CFM stems from the fact that the
correction function is defined as the solution to a PDE,
which, in principle, can be solved to arbitrary order of
accuracy.
In the method described above, the accuracy is determined by
the choice of basis functions used to represent $D\/$ within
each patch.
In this paper, as in ref.~\citep{marques:2011}, we obtain
fourth order of accuracy by representing $D\/$ with Hermite
cubic interpolants.

In addition, the PDE that defines the correction function,
and the least squares procedure used to solve it, do not
depend on the computational grid.
Hence, the CFM is applied exactly the same way for every
discretization stencil that crosses the interface,
and no special cases need to be considered.
This feature makes the CFM very robust with respect to the 
arbitrary fashion an interface can cross the computational
grid.
%

\section{New technique for imposing jump conditions in a
least squares sense}
\label{sec:integrals}

In this section we introduce a new technique for imposing
the jump conditions in a least squares sense.
This technique is based on two main observations:
\begin{enumerate}[(i)]
 \item The jump conditions can be enforced over any
 collection of surfaces that lie within a reasonable
 distance from the center of the correction function patch,
 and whose union has approximately the same area as a
 diagonal plane of the cubic patch.
 \item The level set function can be used to define local
 coordinate transformations that map $L^{\infty}$ balls on
 the interface onto squares.
\end{enumerate}
Hence, we impose the jump conditions on a collection of
$L^{\infty}$ balls that satisfies observation (i) and, per
observation (ii), over which we can easily integrate after
appropriate coordinate transformations.
These coordinate transformations are introduced in
\S\ref{sub:t}.
In order to evaluate surface integrals, we also must be able
to invert the (nonlinear) coordinate transformations
efficiently.
In \S\ref{sub:ti} we present an algorithm to compute
an approximate inverse of these transformations that only
involves solving linear systems of equations.
In \S\ref{sub:ls} we show how to use the coordinate
transformations, and their inverses, to impose jump
conditions in a least squares sense.
Finally, in \S\ref{sub:implementation} we discuss
implementation details.


\subsection{Local coordinate transformation}
\label{sub:t}

\begin{figure}[t!]
 \includegraphics[height=2.2in,
  trim = {0, 0in, 0.5in, 0}, clip]
  {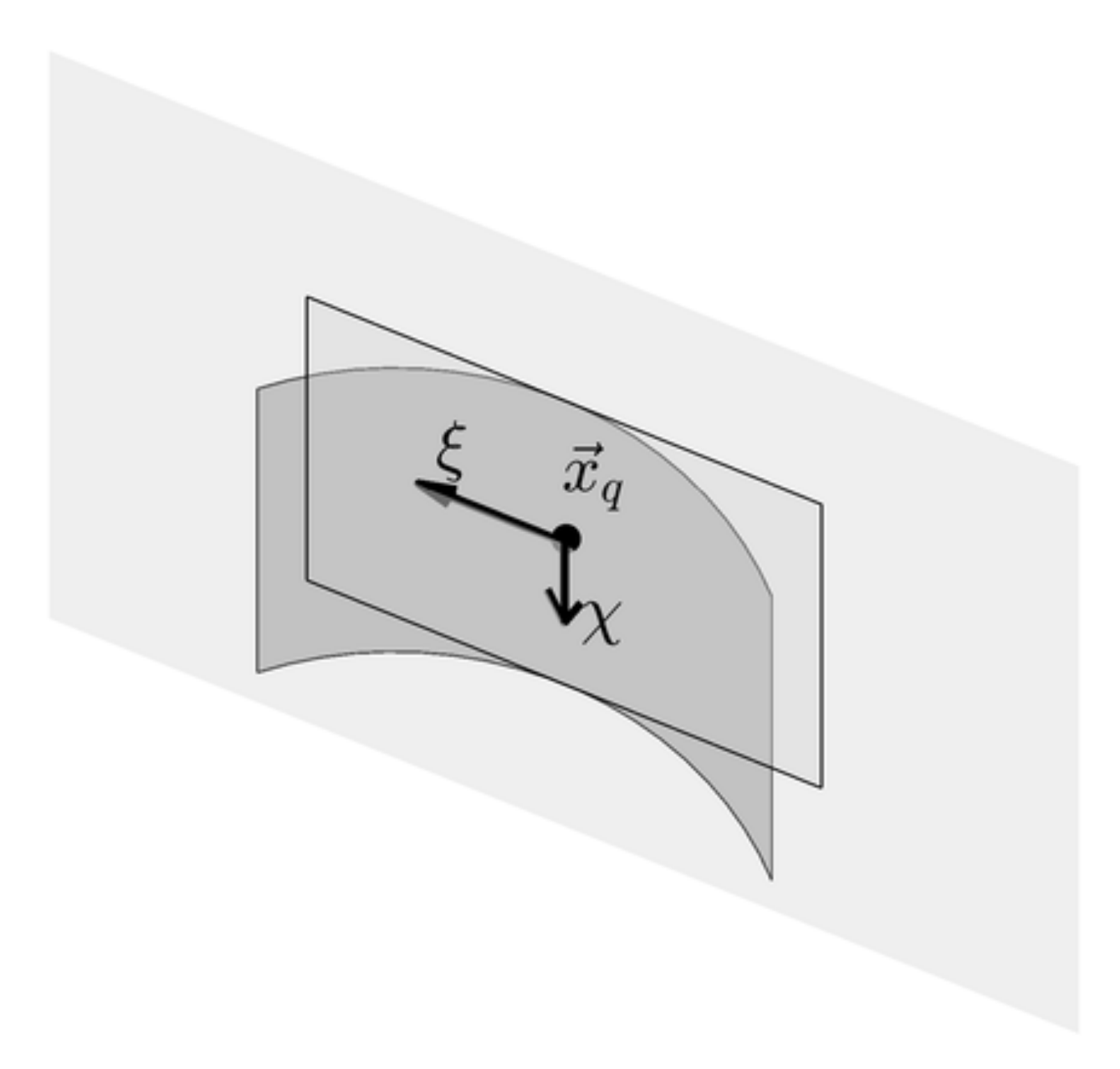}
 \hspace{0.7in}
 \includegraphics[height=2.2in,
 trim = {0.5in, 1.in, 2.5in, 0.3in}, clip]
 {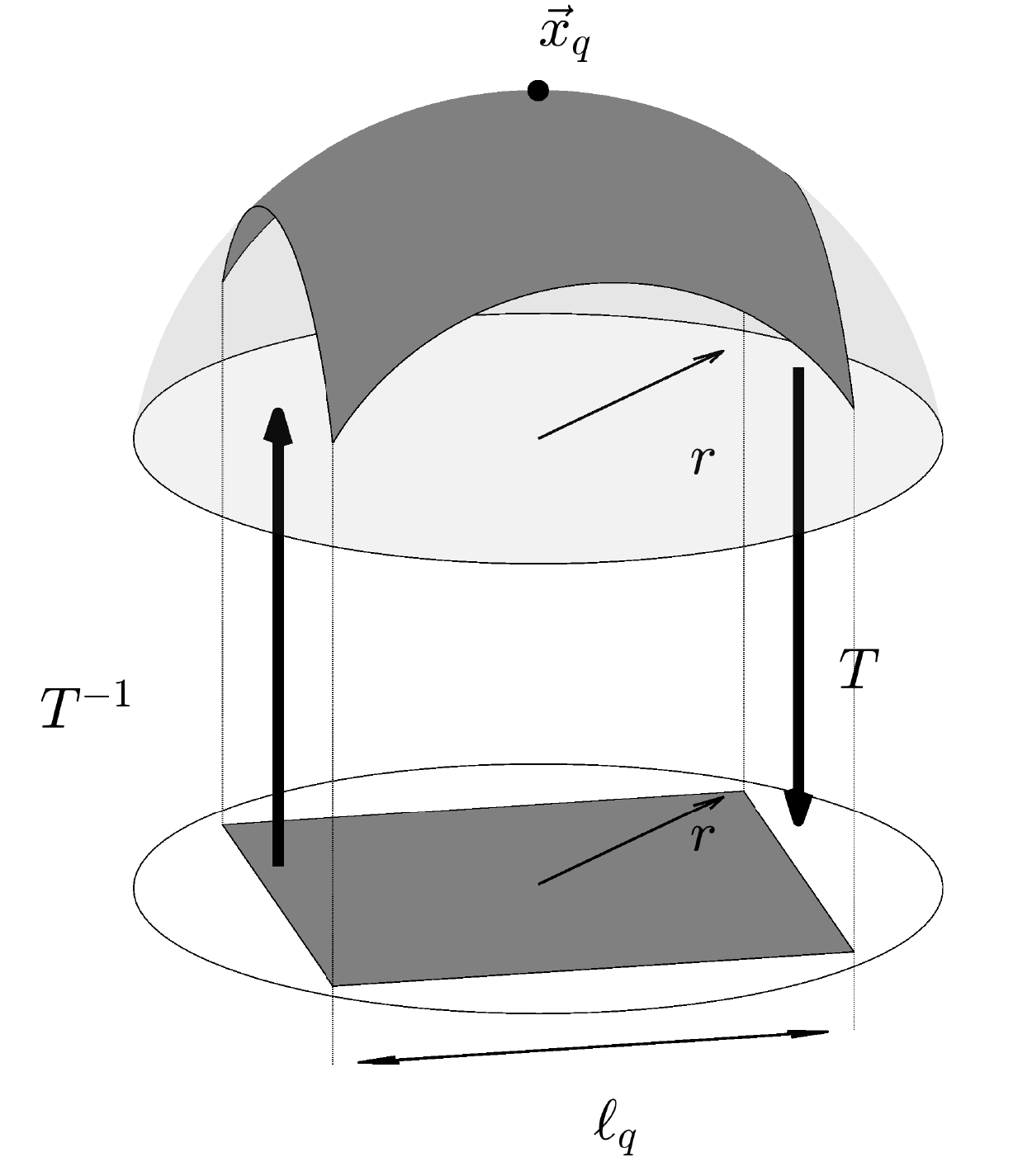}
 \caption{%
 Local coordinate transformation $T_q\/$.
 Left: coordinates $\xi\/$ and $\chi\/$ are defined by the
 orthogonal projection onto the plane tangent to the
 interface at $\x_q\/$.
 Right: spherical interface of radius $r\/$.
 The transformation is bijective if $\ell_q < \sqrt{2}r\/$.
 The dark shaded region denotes \sq\/.
}
 \label{fig:T}
\end{figure}

Let \xq\/ denote a point on the interface $\Gamma\/$.
Then, the following transformation maps a $L^{\infty}$ ball
on the interface centered on \xq\/ onto a square.
This transformation is depicted in figure~\ref{fig:T}(a).

\begin{dfn}\label{dfn:forward}
 \textbf{Local coordinate transformation.}
 Let $\hat{n}\/$ denote the unit vector normal to the
 interface at \xq\/, $\hat{t}_1\/$ and $\hat{t}_2\/$
 denote vectors tangent to the surface, given by
 \eqref{eq:tangent}, and $\phi\/$ denote the level set
 function that represents $\Gamma\/$. 
 Then, the local coordinate transformation
 $\{\eta, \xi, \chi \} = T_q(\x)\/$ is given by
 \begin{equation*}
  \eta = \dfrac{\phi}{||\vec{\nabla}\phi(\xq)||}\/, \quad
  \xi = \hat{t}_1 \cdot (\x - \xq)\/, \quad
  \chi = \hat{t}_2 \cdot (\x - \xq)\/.
 \end{equation*}
 In particular, consider the ball
 \begin{equation*}
  \sq =
  \{ \x \in \Gamma \mid
     ||\x - \xq||_{n_{\infty}} \le \ell_q \}\/,
 \end{equation*}
 where $||(.)||_{n_{\infty}} =
 \max(| (.) \cdot \hat{n} |, \,
 | (.) \cdot \hat{t}_1 |, \,
 | (.) \cdot \hat{t}_2 |)\/$.
 The transformation $T_q\/$ maps \sq\/ onto the square
 $\{\xi, \chi\} \in [-\ell_q/2, \, \ell_q/2]^2\/$ on
 the $\eta = 0\/$ plane.
\end{dfn}

If \sq\/ is close enough to a plane (i.e., $\ell_q\/$ is
much smaller than the minimum radius of curvature), the
transformation is smooth and invertible in \sq\/.
To illustrate this fact, consider the simple case of a
spherical interface of radius $r\/$, as depicted in
figure~\ref{fig:T}(b).
As long as \sq\/ fits strictly within a half-sphere, the
projection is smooth and invertible (because it is for any
spherical cap smaller than a half-sphere).
Simple geometry then shows that, in this case, the criterion
for smoothness and invertibility is $\ell_q < \sqrt{2} r\/$.
For an ellipsoid a similar argument shows that the criterion
is $\ell_q < \sqrt{2} r_{\min}\/$ (note that any surface can
be, locally, approximated by an ellipsoid).
As discussed in \S\ref{sub:ti}, the algorithm used to
approximate the inverse transformation $T_q^{-1}\/$ imposes
a more stringent restriction to $\ell_q\/$.

In practice, we link $\ell_q\/$ to the size of the
computational grid used to represent the level set function,
and thus is set by the user.
As described in~\S\ref{sub:implementation}, our
implementation uses a convenient heuristics to check if the
grid provided by the user has adequate refinement to satisfy
the restrictions on $\ell_q\/$.

In addition, \xq\/ need not lie exactly on the interface.
If \xq\/ is close to the interface (in the
$\ell_q\/$-scale), then $\xi\/$ and $\chi\/$ are given by
an orthogonal projection onto a plane approximately tangent
to the interface.
Hence $T_q\/$ is well defined, and the arguments above
about smoothness and invertibility still apply.

\subsection{Inverse transformation}
\label{sub:ti}

Here we introduce an algorithm that computes an
approximation to the inverse transformation $T_q^{-1}\/$.
This algorithm avoids having to solve a nonlinear system of
equations whenever an evaluation of $T_q^{-1}\/$ is required,
which leads to significant computational savings.

\begin{figure}[t!]
 \begin{center}
  \includegraphics[scale=0.5,
   trim = {1.5in, 1in, 1.5in, 0.5in}, clip]
   {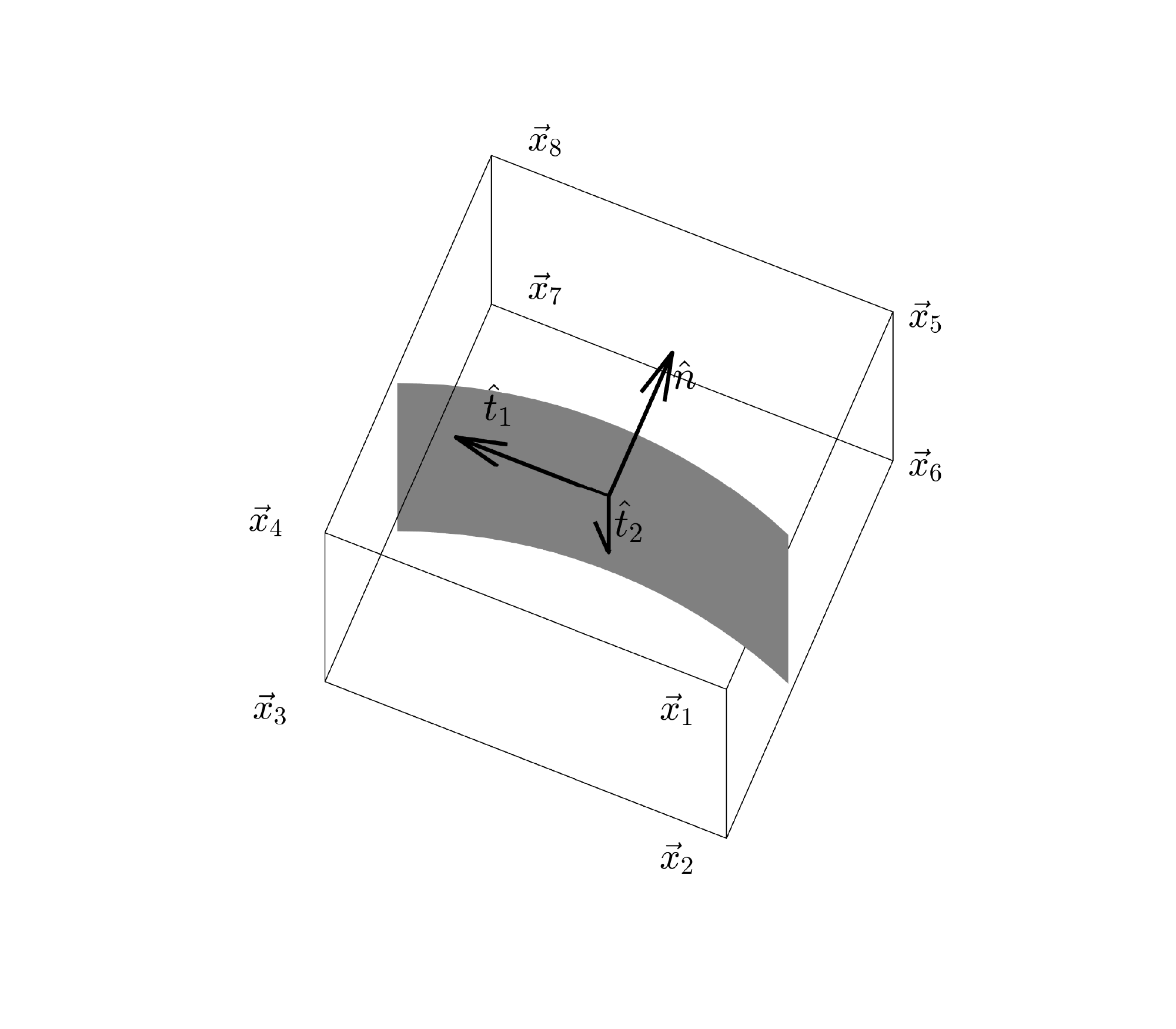}
 \end{center}
 \vspace{-0.2in}
 \caption{Cube \cb\/ used to compute an approximation to
          the inverse transformation $T_q^{-1}\/$.}
 \label{fig:cube}
\end{figure}

Assume that the level set function $\phi\/$ is represented
by a polynomial of degree $p\/$.
Let $\hat{n}\/$ denote the unit vector normal to interface
at \xq\/, and $\hat{t}_1\/$ and $\hat{t}_2\/$ denote vectors
that complete an orthonormal basis computed
with~\eqref{eq:tangent}.
Then, we define \cb\/ as a cube of edge length $\ell_q\/$
centered at \xq\/, as follows,
\begin{equation*}
 \cb =
  \left\{ \x \; \bigg \vert \;
     \x = \xq
        + \dfrac{\ell_q}{2}
        ( \gamma_1 \hat{t}_1 + \gamma_2 \hat{t}_2 
        + \gamma_3 \hat{n} ), 
     \, \gamma_i \in [-1, 1] \right\}.
\end{equation*}
Figure~\ref{fig:cube} depicts the cube \cb\/.
Algorithm~\ref{alg:ti} computes an approximation to
$T_q^{-1}\/$ within the region $T_q(\cb)\/$ using a
Hermite polynomial of degree $p\/$.

\begin{algorithm}
 \caption{Approximate inverse transformation}
 \label{alg:ti}
 \begin{algorithmic}[1]
  \vspace{0.03in}
  \State Compute $\theta_i = T_q(\x_i)\/$, where $\x_i\/$,
         $i = 1, \ldots, 8\/$ denotes the 8 vertices of
         \cb\/.
  \vspace{0.05in}
  \State Compute all derivatives
         $\partial^{|\alpha|} T_q(\x_i) /
          \partial x^{\alpha_1} \partial y^{\alpha_2}
          \partial z^{\alpha_3}\/$, with
          $|\alpha| = \sum \alpha_i$, and
          $\max(\alpha_i) \le p-2\/$.
  \vspace{0.05in}
  \State Using the inverse function theorem, compute all
         derivatives
         $\partial^{|\alpha|} T_q^{-1}(\theta_i) /
          \partial \eta^{\alpha_1} \partial \xi^{\alpha_2}
          \partial \chi^{\alpha_3}\/$, with
          $\max(\alpha_i) \le p-2\/$.
  \vspace{0.05in}
  \State Solve for the Hermite polynomial of degree $p\/$
         that fits the inverse transformation at $\theta_i\/$.
 \end{algorithmic}
\end{algorithm}

The accuracy of the approximation to $T^{-1}\/$ imposes an
upper bound on $\ell_q\/$. 
The Hermite interpolation results in an error
$\ord{\ell_q^{p+1}D^{(p+1)}T^{-1}}\/$, where $D^{(p+1)}\/$
denotes derivatives of order $p+1\/$.
Hence, an accurate approximation is only possible when
derivatives of order $p+1\/$ are bounded.
Furthermore, the derivatives of $T^{-1}\/$ are related to
the ratio $\ell_q/r_{\min}\/$, i.e., to how flat the section
of the interface is with respect to $\ell_q\/$.
We conjecture that, for a given $p\/$, it is possible to
choose a constant $c > 0\/$ such that
$\ell_q < c \, r_{\min}\/$ guarantees
$D^{(p+1)}T^{-1} = \ord{1}\/$.
In particular, in our implementation we use $p=3\/$.
For a spherical cap section of the interface, and $p = 3\/$,
one can show that $c \approx 0.4\/$.
This observation motivates the heuristics we use to
determine an upper bound for $\ell_q\/$, as discussed in
\S\ref{sub:implementation}.


\subsection{Imposing jump conditions}
\label{sub:ls}

Our goal is to impose jump conditions over a piece of the
interface that includes enough information to define a
unique solution to the Caucy problem~\eqref{eq:D} within
each patch $\Omega_{\Gamma_i}\/$.
In ref.~\citep{marques:2011} we show that in practice the
method works well if these conditions are imposed over a
piece of the interface whose area approximates the area of a
diagonal plane of the correction function patch, and that is
located within a ball of radius $\ord{\ell_i}$ centered at
the patch.
There are many choices of surfaces that satisfy this
heuristics.
Previous implementations of the CFM impose jump conditions
on the intersection between the interface and the correction
function patch.
Here we use the local transformation in \S\ref{sub:t} to
define the section of interface over which the jump
conditions are imposed, which simplifies the computation of
surface integrals via numerical quadrature.

Given $\xq \in \Gamma\/$, the corresponding local
transformation $T_q\/$, and its inverse $T_q^{-1}\/$, one
can evaluate the square error in satisfying the jump
conditions over the surface \sq\/ as follows:

\begin{equation}\label{eq:int}
 \begin{split}
  \int_{\sq} \mathscr{E} \, dS
  &=
  \int_{-\ell_s/2}^{\ell_s/2}
  \int_{-\ell_s/2}^{\ell_s/2}
  \mathscr{E}(T_q^{-1}\bigl(0, \xi, \chi)\bigr)
  \biggl\lVert
  \dfrac{\partial\vec{x}}{\partial\xi}\times
  \dfrac{\partial\vec{x}}{\partial\chi}
  \biggr\rVert \, d\xi\, d\chi\/ \\[5pt]
  &\approx
  \sum_j
   \Biggl( w_j \,
    \mathscr{E}(T_q^{-1}\bigl(0, \xi_j, \chi_j)\bigr)
   \biggl\lVert
   \dfrac{\partial\vec{x}}{\partial\xi}\times
   \dfrac{\partial\vec{x}}{\partial\chi}
   \biggr\rVert_{(0, \xi_j, \chi_j)}\Biggr)\/,
 \end{split}
\end{equation}
where
\begin{equation*} 
 \mathscr{E} = (D - a)^2 + \ell_i^2(D_n - b)^2\/,
\end{equation*}
$w_j\/$ are the weights of a numerical quadrature defined
over the square \linebreak
$[-\ell_q/2, \ell_q/2]^2\/$, and $(\xi_j, \chi_j)\/$ are the
quadrature points.

\begin{figure}[t!] 
 \begin{center}
  \includegraphics[scale=0.6]{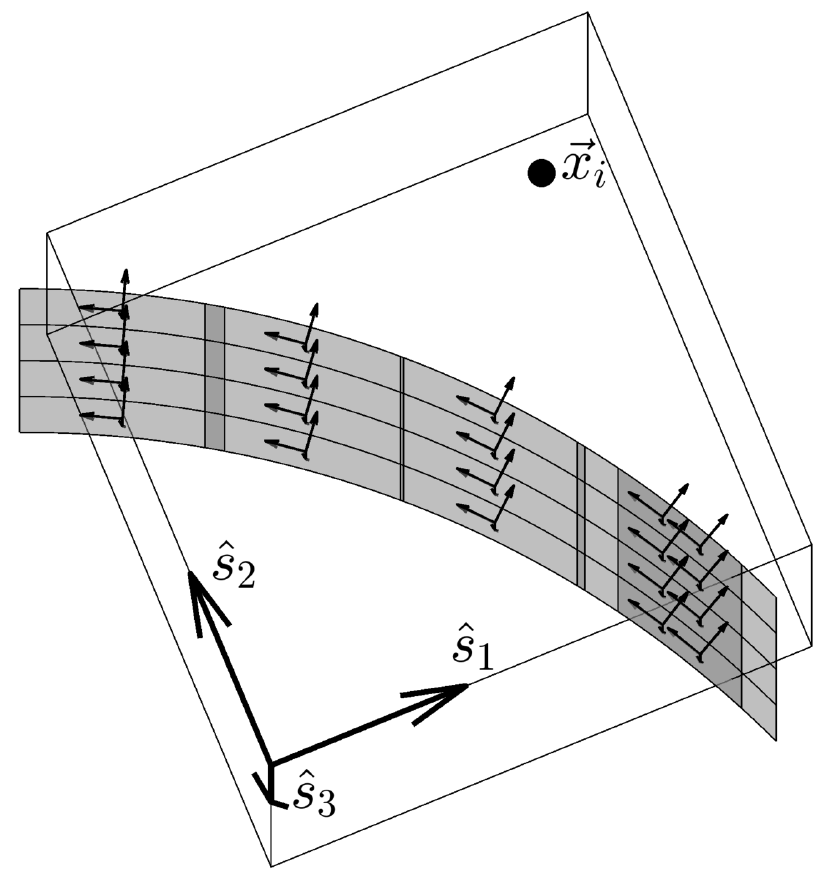}\\
 \end{center}
 \vspace{-0.2in}
 \caption{The surface integrals are evaluated over
          $L^{\infty}$ balls over the interface.
          The sections may overlap, and one patch may
          contain several sections.}
 \label{fig:sections}
\end{figure}

As discussed in \S\ref{sub:implementation}, the location of
the points $\x_q\/$ and the length $\ell_q\/$ are defined
by the computational grid used to represent the level set
function.
In general, the patch $\Omega_{\Gamma_i}\/$ can intersect
several sections \sq\/, each associated with a different
\xq\/, as illustrated by figure~\ref{fig:sections}.
Hence, within each patch we impose the jump conditions over
the union of all \sq\/ regions associated with the set
$Q = \{ q \mid
||\x_q - \x_{0_i}||_2 < \sqrt{2}/2 \, \ell_i\/ \}\/$.
We incorporate this new technique of imposing jump
conditions into the CFM by modifying the minimization
functional~\eqref{eq:J} to
\begin{equation}\label{eq:jnew}
 \begin{split}
  J_i^{\text{new}} &=
  \dfrac{\ell_i^4}{V_i}
  \int_{\Omega_{\Gamma_i}}
  (\nabla^2 D - f_D)^2 dV\\
  &+   \dfrac{1}{\tilde{S}_i} \sum_{q \in Q}
  \int_{\sq}
  [(D-a)^2 + \ell_i^2(D_n-b)^2] dS\/,
 \end{split}
\end{equation}
where $\tilde{S}_i = \sum_{q \in Q} S_q\/$, and $S_q\/$
denotes the area of \sq\/.

\subsection{Implementation details}
\label{sub:implementation}

We use the discretization of the level set function $\phi\/$
to compute the set of points $\x_q\/$ and to determine the
length $\ell_q\/$.
Let $G\g\/$ denote the grid used to discretize $\phi\/$,
and $\xg\/$ denote the set of grid cells in
$G\g\/$ that are crossed by the interface.
Then, for each cell $q \in \xg$ we set $\ell_q\/$ as the
diagonal length of the cell, and
$\x_q = P_{\Gamma}(\x_{c_q})\/$, where $\x_{c_q}\/$ denotes
the center of the cell.
This implementation assumes that $G\g\/$ is fine enough to
``resolve'' the interface $\Gamma\/$, such that the local
transformation $T_q\/$ is smooth and invertible, and that
the approximate algorithm used to compute $T_q^{-1}\/$ is
accurate.

In practice, we assess the adequateness of $\ell_q\/$ by
monitoring the determinant of the Jacobian on the
$\xi-\chi\/$ plane,
\begin{equation}\label{eq:jac}
 \text{Jac} =
 \biggl\lVert
 \dfrac{\partial\vec{x}}{\partial\xi}\times
 \dfrac{\partial\vec{x}}{\partial\chi}
 \biggr\rVert_{T^{-1}(0,\xi,\chi)}\/,
\end{equation}
evaluated at the quadrature points used to compute the
surface integral~\eqref{eq:int}.
Since~\eqref{eq:jac} is computed to impose the jump
conditions, computing this indicator does not add to the
computational cost.
For a spherical cap section of the interface we can show
that $\ell_q < 0.4 \, r\/$ implies $\text{Jac} > 0.83\/$
(see \S\ref{sub:ti} for the motivation for this limitation
on $\ell_q\/$).
Hence, we use this threshold to verify whether $\ell_q\/$ is
small enough in general cases.

Furthermore, to guarantee that $G\g\/$ can always be made
as fine as needed, we allow $G\g\/$ to be distinct from the
grid used to solve the Poisson equation.
We make this distinction explicit by denoting the grid in
which the Poisson equation is solved by $G_P\/$.


\section{Results} \label{sec:results}
%
In this section we verify the accuracy and robustness of the
proposed technique for imposing jump conditions with
numerical examples in 2-D and 3-D.

The results shown here are computed with an overall fourth
order accurate scheme.
Namely, we discretize Poisson's equation using the standard
9-point stencil in 2-D, and the equivalent 19-point stencil
in 3-D.
Furthermore, we represent the correction function with
Hermite cubic polynomials, which is consistent with fourth
order of accuracy.
Finally, we represent the interfaces using the
Gradient-Augmented Level-Set (GALS)
method~\citep{nave:2010}, which is also based on Hermite
cubic polynomials.

Below we use analytic expressions to define the model
problems, but only the appropriate discrete data defined on
a computational grid are used as inputs to the code (this
is close to a practical situation in which data defining the
interface is the result of a computation on the same grid).
The data is presented in terms of the exact solution
(denoted by $u\/$), and the level set function (denoted by
$\phi\/$).
We verify accuracy by evaluating the $L^{\infty}\/$ norm of
the error in the solution and its gradient.
Furthermore, in some problems the resolutions of $G_P\/$
(grid where Poisson's equation is solved) and $G_{\Gamma\/}$
(grid where the level set function is defined) are distinct.
In these cases we maintain the resolution ratio between
these grids constant during error convergence studies.

In the first two examples we compare the accuracy of the
technique presented here to the approach used in
ref.~\citep{marques:2011} to solve 2-D problems.
The CFM implemented in ref.~\citep{marques:2011} differs
from the current work in three main aspects:
  (i)~the patch used to compute the correction function,
 (ii)~the weight given to the jump condition in the least
      squares
      minimization~(\ref{eq:ls_minimization}--\ref{eq:J}),
      and
(iii)~the technique used to compute the integrals along the
      interface.
Since the focus of the present paper is point (iii), in
these examples we compute two sets of solutions:
minimizing \eqref{eq:jnew} with the integration technique of
\S\ref{sec:integrals} (coordinate transformation), and
minimizing \eqref{eq:J} with the integration technique of
ref.~\citep{marques:2011} (curve parametrization).
Both solutions are based on the patch construction described
in \S\ref{sub:patch}.
We compare these solutions with the results presented in
ref.~\citep{marques:2011}.

\subsection{Smooth 5-pointed star} \label{sub:results:star}
%
Here we reproduce example 2 of ref.~\citep{marques:2011},
which involves a smooth 5-pointed star.
By considering a smooth interface we guarantee that the
transformation introduced in \S\ref{sub:t} is well defined,
without the need of special considerations near singular
points (e.g., corners).
In this example $G_P$ and $G\g$ are the same grid.
The problem is defined as follows.
%
\begin{itemize}
 \item $\phi(x\/,\,y) = (x-0.5)^2 + (y-0.5)^2 -
        \big(0.25 + 0.05\sin(5\varphi(x\/,\,y)\big)^2\/$.
 \item $\varphi(x\/,\,y) =
        \arctan\Big(\dfrac{y-0.5}{x-0.5}\Big)\/$.
 \item $\Omega_1 = \{(x\/,\,y) \in [0,1]^2 \mid
        \phi(x\/,\,y) \le 0\}\/$.
 \item $\Omega_2 = \{(x\/,\,y) \in [0,1]^2 \mid
        \phi(x\/,\,y) > 0\}\/$.
 \item $u_1(x\/,\,y) = \exp(x)\cos(y)\/$.
 \item $u_2(x\/,\,y) = 0\/$.
\end{itemize}

\begin{figure}[t!] 
 \begin{center}
  \includegraphics[height = 2.1in]{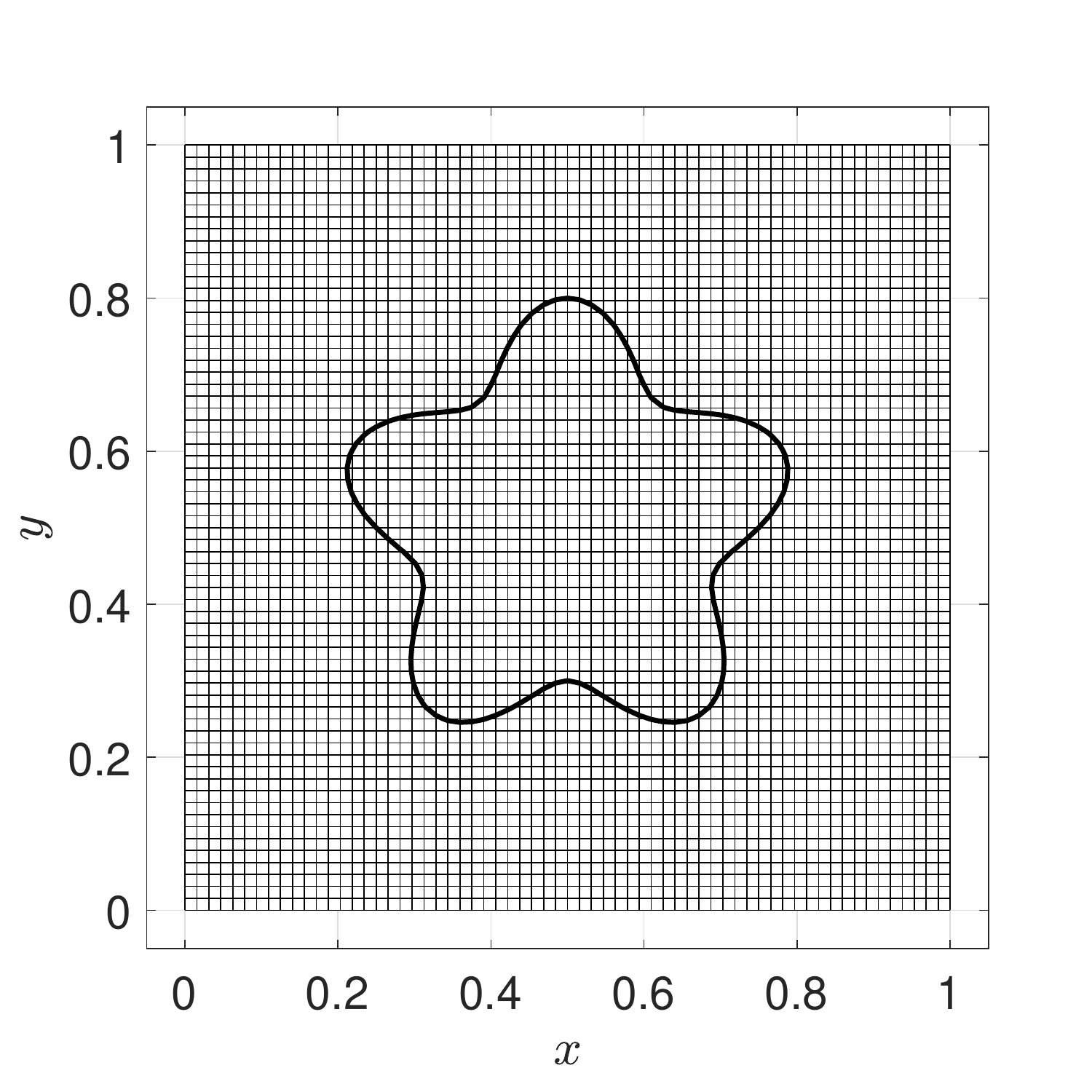}
  \hfill
  \includegraphics[height = 2.1in]{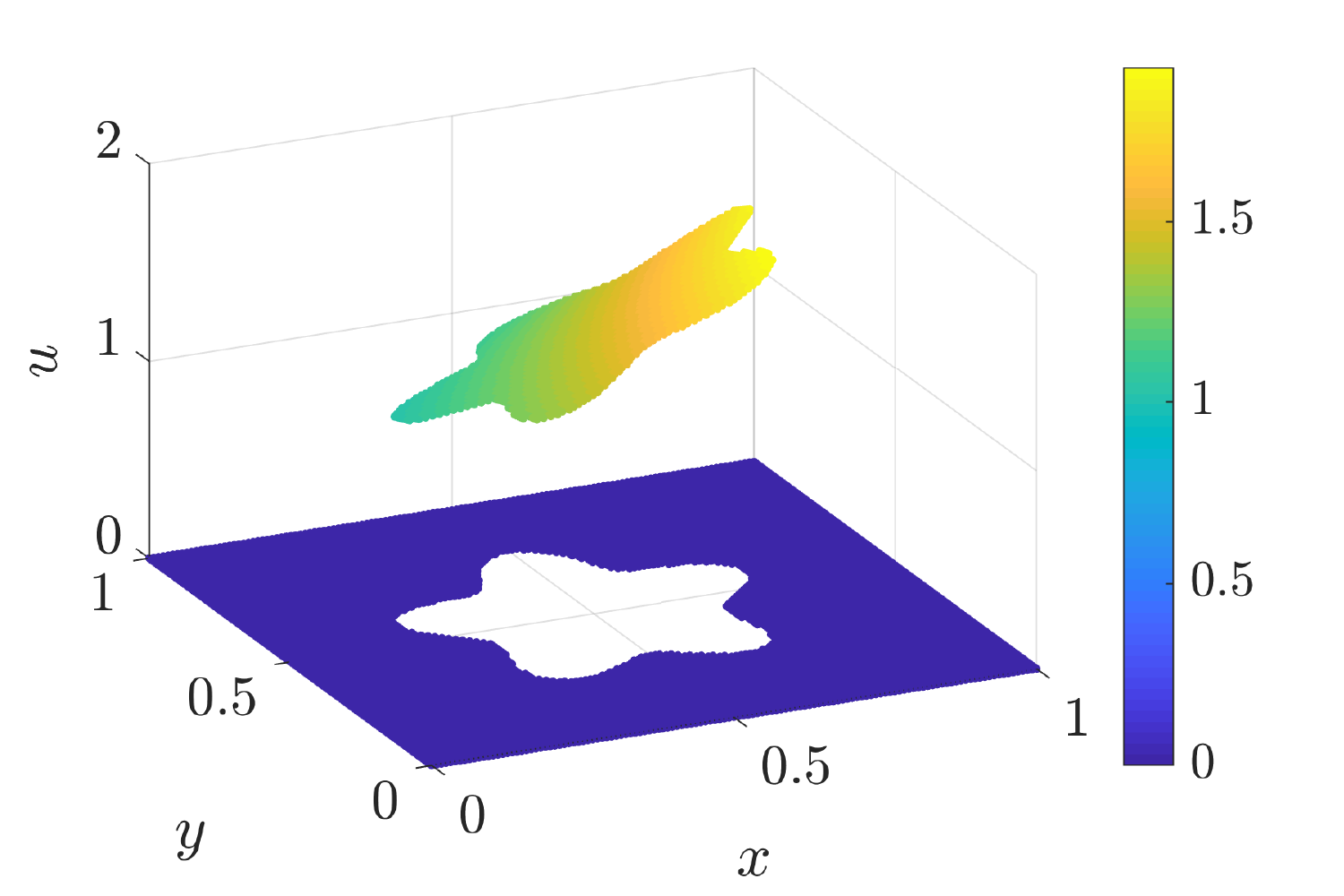}
 \end{center}
 \vspace{-0.2in}
 \caption{Example 1.
          Left: interface immersed into $G_P\/$.
          Right: Solution given by the CFM.}
 \label{fig:star:grid}
\end{figure}

\begin{figure}[t!] 
 \begin{center}
  \includegraphics[width = 3.5in]{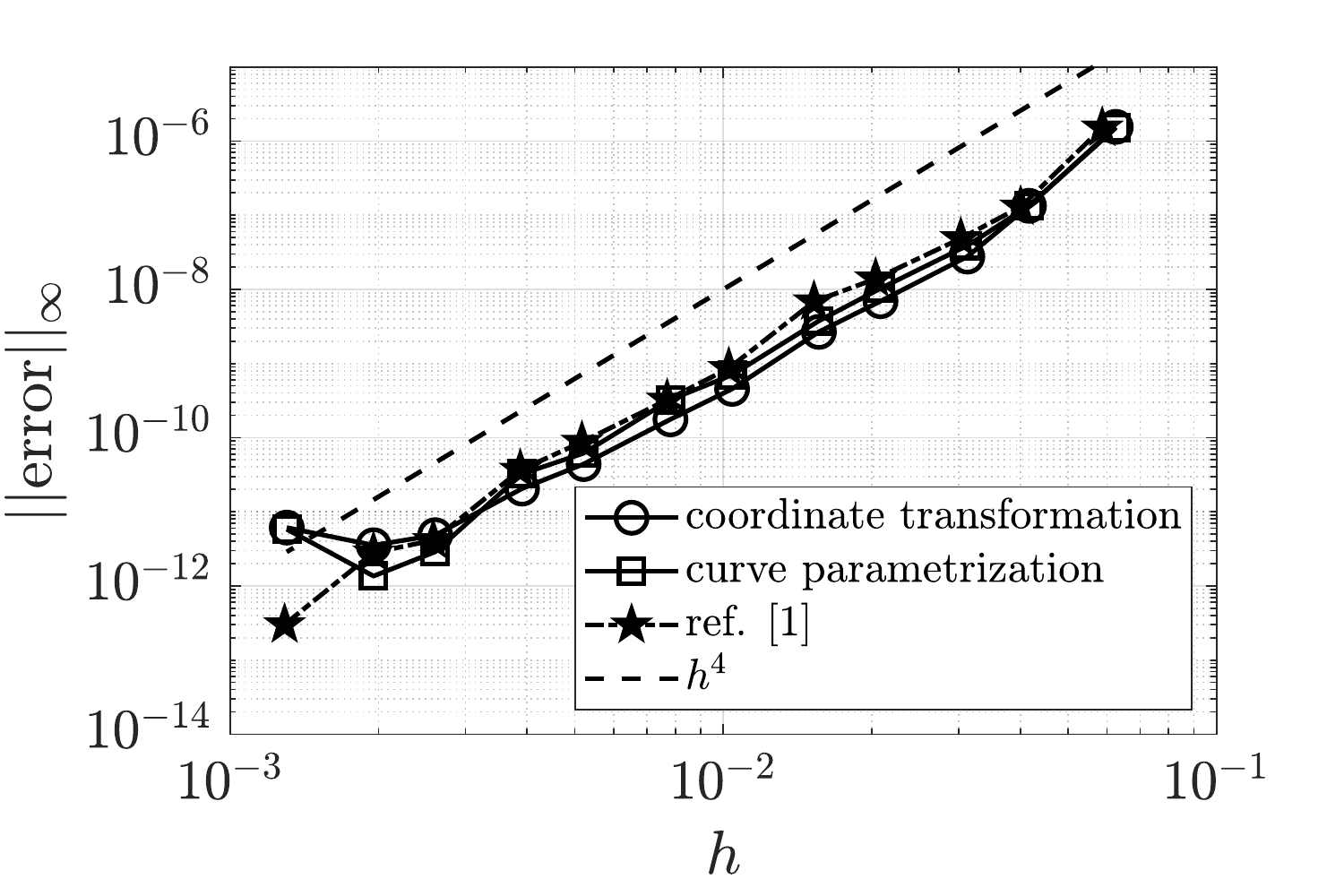}
 \end{center}
 \vspace{-0.2in}
 \caption{Example 1.
          Error convergence in the $L^{\infty}\/$ norm.
          Comparison of results computed by evaluating
          surface integrals with coordinate transformation
          (current work), curve parametrization, and
          results presented in ref.~\citep{marques:2011}.}
 \label{fig:star:error}
\end{figure}

Figure~\ref{fig:star:grid} shows the interface immersed into
the Cartesian grid $G_P$ used to solve the Poisson equation,
along with a plot of the solution obtained with the CFM.
Figure~\ref{fig:star:error} shows the error obtained with
the present technique (coordinate transformation), the
current version of the CFM with integration technique of
ref.~\citep{marques:2011} (curve parametrization), and the
results presented in ref.~\citep{marques:2011}.
All three versions of the CFM present fourth order of
accuracy and comparable errors.

\subsection{Touching circles} \label{sub:results:circles}
%
Here we reproduce example 5 of ref.~\citep{marques:2011},
which involves two circular interfaces that touch at one
point.
In this example the interfaces are represented by level set
functions defined on independent grids: $G\gn{1}\/$ and
$G\gn{2}\/$.
To be consistent with the implementation of
ref.~\citep{marques:2011}, we choose $G\gn{1}\/$ and
$G\gn{2}\/$ to be the same as $G_P\/$.
The problem is defined as follows.

\begin{itemize}
 \item $\phi_1(x\/,\,y) = \big(x-0.5-0.2\cos(\pi /e^2)\big)^2 +
        \big(y-0.5-0.2\sin(\pi /e^2)\big)^2 - 0.01\/$.
 \item $\phi_2(x\/,\,y) = (x-0.5)^2 + (y-0.5)^2 - 0.09\/$.
 \item $\Omega_1 = \{(x\/,\,y) \in [0,1]^2 \mid
        \phi_1(x\/,\,y) \le 0\}$.
 \item $\Omega_2 = \{(x\/,\,y) \in [0,1]^2 \mid
        \phi_1(x\/,\,y) > 0, \phi_2(x\/,\,y) \le 0\}\/$.
 \item $\Omega_3 = \{(x\/,\,y) \in [0,1]^2 \mid
        \phi_1(x\/,\,y) > 0, \phi_2(x\/,\,y) > 0\}\/$.
 \item $u_1(x\/,\,y) = \sin(\pi x)\sin(\pi y) + 5\/$.
 \item $u_2(x\/,\,y) = \sin(\pi x)\big(\sin(\pi y)
        - \exp(\pi y)\big)\/$.
 \item $u_3(x\/,\,y) = \exp(x)\big(x^2\sin(y) + y^2\big)\/$.
\end{itemize}

\begin{figure}[t!] 
 \begin{center}
  \includegraphics[height = 2.1in]{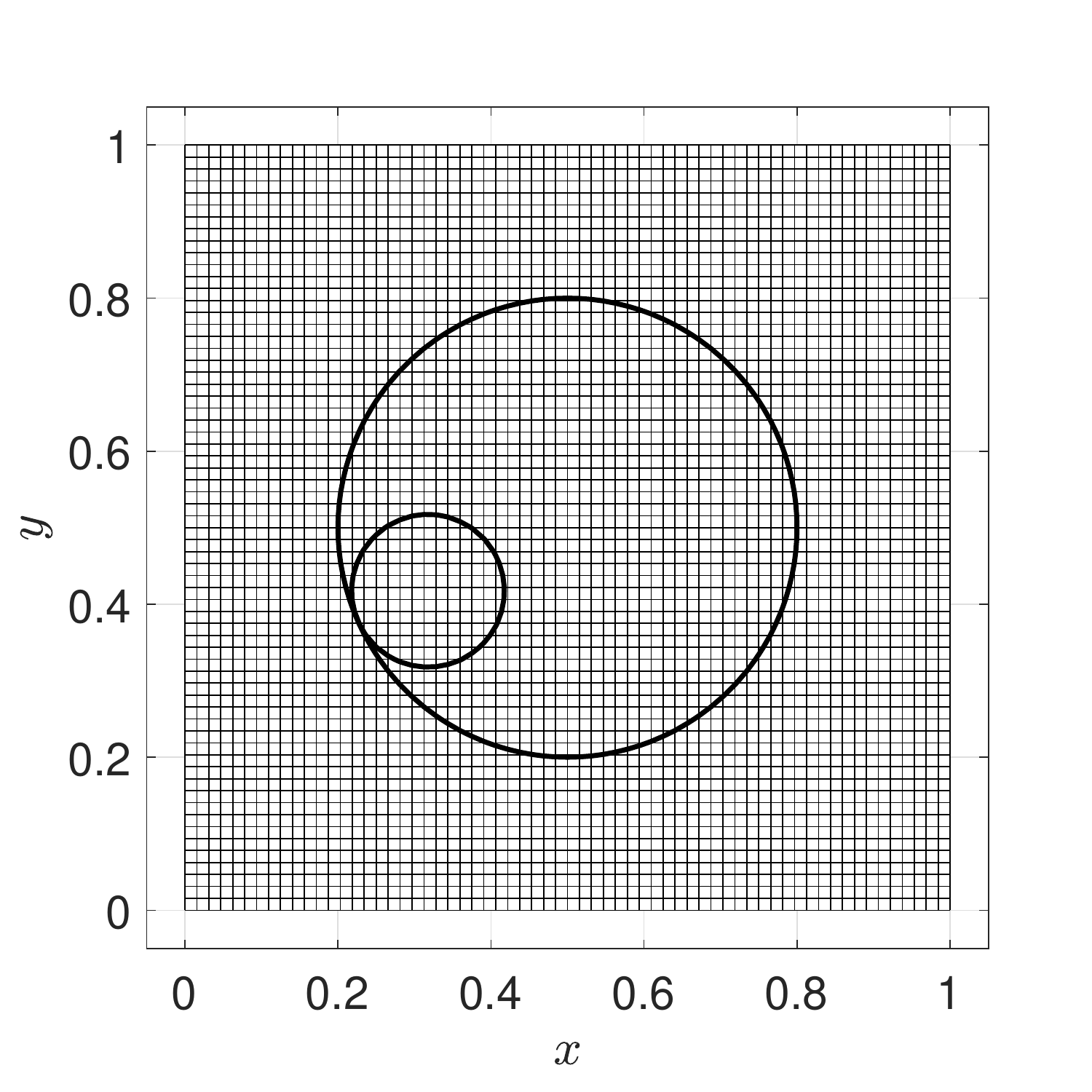}
  \hfill
  \includegraphics[height = 2.1in]{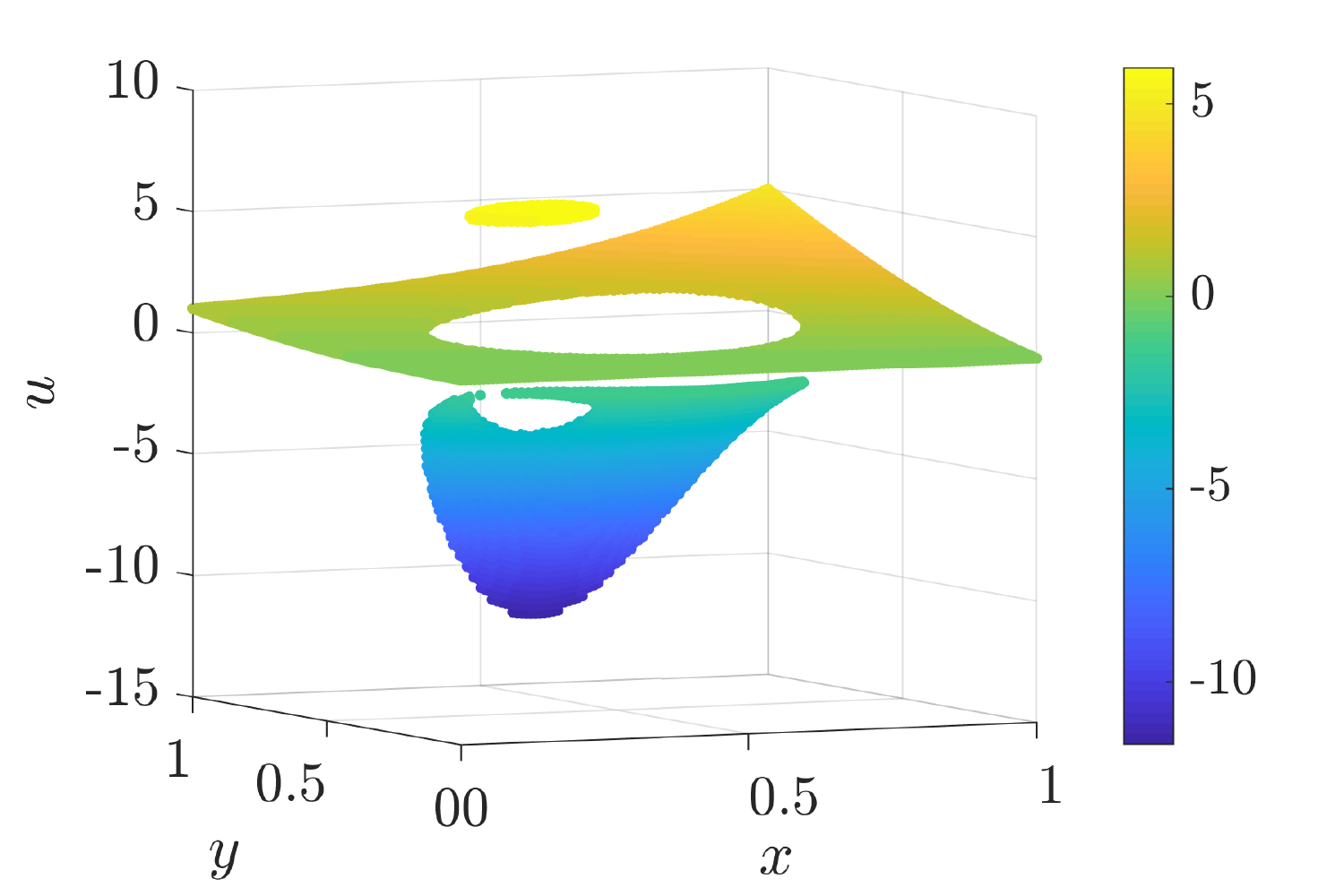}\\
 \end{center}
 \vspace{-0.2in}
 \caption{Example 2.
          Left: interfaces immersed into $G_P\/$, with
          circles that touch at single point.
          Right: solution obtained with the CFM.}
 \label{fig:circles:grid}
\end{figure}

Figure~\ref{fig:circles:grid} shows the interfaces immersed
into the Cartesian grid $G_P$ used to solve the Poisson
equation, along with a plot of the solution obtained with
the CFM.
Since the jump conditions (\ref{eq:a}-c) are linear, one can
solve for correction functions associated with each
interface independently and combine them as needed -- see
remark~\ref{rmk:multiple_int}.
As shown in fig.~\ref{fig:circles:grid}, this approach
allows for arbitrarily close interfaces

\begin{rmk}\label{rmk:multiple_int}
 In practice, the interfaces seen by the CFM are subject to
 errors due to the level set representation. Hence, the
 ``effective'' interfaces likely do not touch perfectly at
 just one point.
 They may cross over, or not touch at all.
 However, the CFM can handle these situations seamlessly by
 computing correction functions due to each of the
 interfaces independently.
 For instance, when the solution domain is subdivided into
 three regions by two interfaces, such as in
 figure~\ref{fig:circles:grid}, the CFM computes
 $D_{12} = u_2 - u_1\/$ and ${D_{13} = u_3 - u_1}\/$, where
 $u_i\/$ denotes the solution restricted to each of the
 three regions.
 Since the
 jump conditions (\ref{eq:a}-c) are linear, these correction
 functions can be combined to compute
 ${D_{23} = u_3 - u_2 = D_{13} - D_{12}}\/$ as
 needed.
 \myremarkend
\end{rmk}

\begin{figure}[t!] 
 \begin{center}
  \includegraphics[width = 3.5in]
   {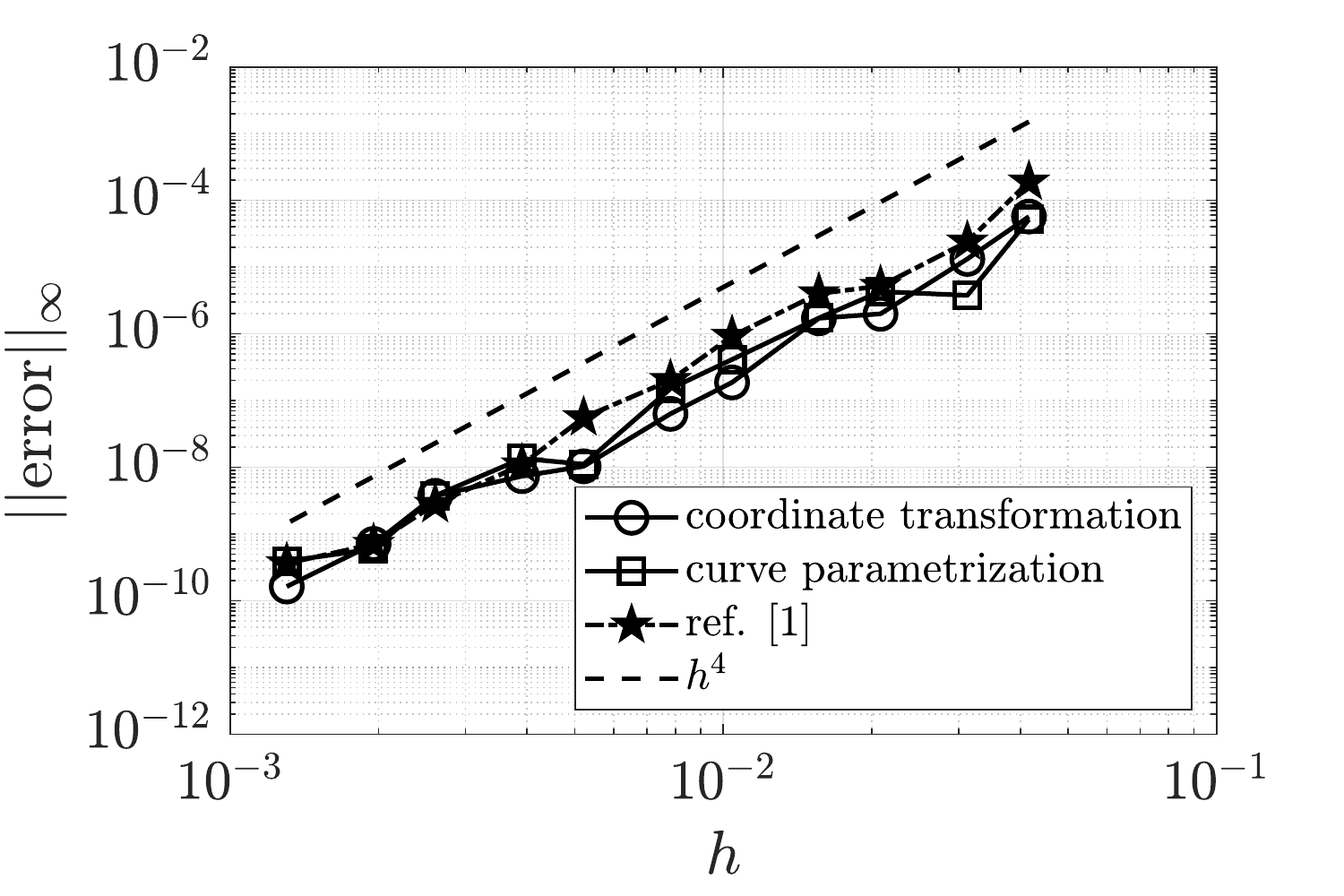}
 \end{center}
 \vspace{-0.2in}
 \caption{Example 2.
          Error convergence in the $L^{\infty}\/$ norm.
          Comparison of results computed by evaluating
          surface integrals with coordinate transformation
          (current work), curve parametrization, and
          results presented in ref.~\citep{marques:2011}.}
 \label{fig:circles:error}
\end{figure}

Figure~\ref{fig:circles:error} shows the error obtained with
the present technique (coordinate transformation), the
current version of the CFM with integration technique of
ref.~\citep{marques:2011} (curve parametrization), and the
results presented in ref.~\citep{marques:2011}.
Once again, all three versions of the CFM present fourth
order of accuracy and comparable errors.

\subsection{Three interfaces in 2-D} \label{sub:results:3int}
%
This 2-D example, which is not in \citep{marques:2011},
illustrates the application of the new implementation of the
CFM to solve problems with more than two interfaces close
together.
We solve the following problem with three interfaces:

\begin{itemize}
 \item $\phi_1(x_1\/,\,y_1) = (x_1-0.3)^2 + (y_1-0.3)^2 -
        0.05\/$.
 \item $\phi_2(x_2\/,\,y_2) = (x_2-0.3)^2 + (y_2-0.3)^2 -
        \bigl(\sqrt{3}/10 +
             0.05\sin(5\varphi_2(x_2\/,\,y_2)\bigr)^2\/$.
 \item $\phi_2(x_3\/,\,y_3) = (x_3-0.25)^2 + (y_3-0.25)^2 -
        \big(0.15 + 0.05\sin(2\varphi_3(x_3\/,\,y_3)\big)^2\/$.
 \item $\varphi_2(x_2\/,\,y_2) =
        \arctan\Big(\dfrac{y_2-0.3}{x_2-0.3}\Big)\/$.
 \item $\varphi_3(x_3\/,\,y_3) =
        \arctan\Big(\dfrac{y_3-0.25}{x_3-0.25}\Big)\/$.
 \item $\Omega_1 = \{(x_1\/,\,y_1) \in [0,0.6]^2 \mid
        \phi_1(x_1\/,\,y_1) \le 0\}\/$.
 \item $\Omega_2 = \{(x_2\/,\,y_2) \in [0,0.6]^2 \mid
        \phi_2(x_2\/,\,y_2) \le 0\}\/$.
 \item $\Omega_3 = \{(x_3\/,\,y_3) \in [0,0.5]^2 \mid
        \phi_3(x_3\/,\,y_3) \le 0\}\/$.
 \item $\Omega_4 = \{(x\/,\,y) \in [0,1]^2 \,|\,
        \phi_i(x\/,\,y) > 0, i = 1\/,\,2\/,\,3\}\/$.
 \item $u_1(x\/,\,y) = \exp(x)\big(x^2\sin(y) + y^2\big)\/$.
 \item $u_2(x\/,\,y) = \sin(\pi x)\sin(\pi y) + 10\/$.
 \item $u_3(x\/,\,y) = x\,y + 10\/$.
 \item $u_4(x\/,\,y) = 10(x^2 + y^2)\/$.
\end{itemize}

In this example the level set grids ($G\gn{1}\/$,
$G\gn{2}\/$, and $G\gn{3}\/$) are not the same as $G_P\/$.
The relationships between the grid spacings are
$h\gn{1}/h_P = h\gn{2}/h_P = 0.6$ and $h\gn{3}/h_P = 0.5$.
In the expressions above, each level set is defined in terms
of the Cartesian coordinates aligned with the corresponding
grid. These coordinates are defined as follows:
\begin{subequations}\label{eq:3int:xi}
 \begin{align*}
  x_1 &= x - 0.34,\\
  y_1 &= y - 0.37,\\
  x_2 &= (x - 0.29)\cos(35\pi /180)
       + (y + 0.1)\sin(35\pi /180)\/,\\
  y_2 &= -(x - 0.29)\sin(35\pi /180) 
       + (y + 0.1)\cos(35\pi /180)\/,\\
  x_3 &= (x - 0.885)\cos(75\pi /180)
       + (y + 0.048)\sin(75\pi /180)\/,\\
  y_3 &= -(x - 0.885)\sin(75\pi /180)
       + (y + 0.048)\cos(75\pi /180)\/.
 \end{align*}
\end{subequations}

\begin{figure}[t!] 
 \begin{center}
  \includegraphics[height = 2.1in]{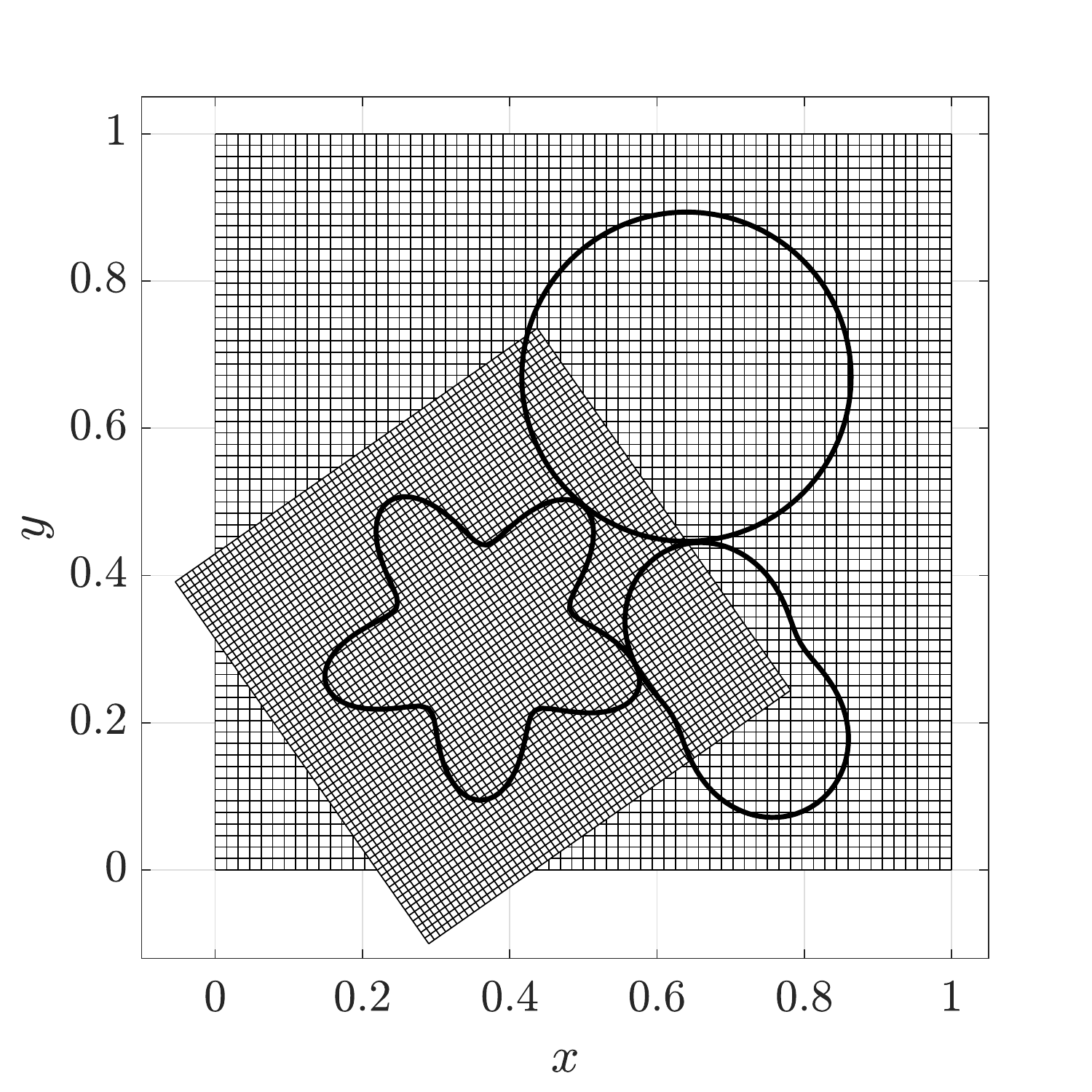}
   \hfill
  \includegraphics[height = 2.1in]{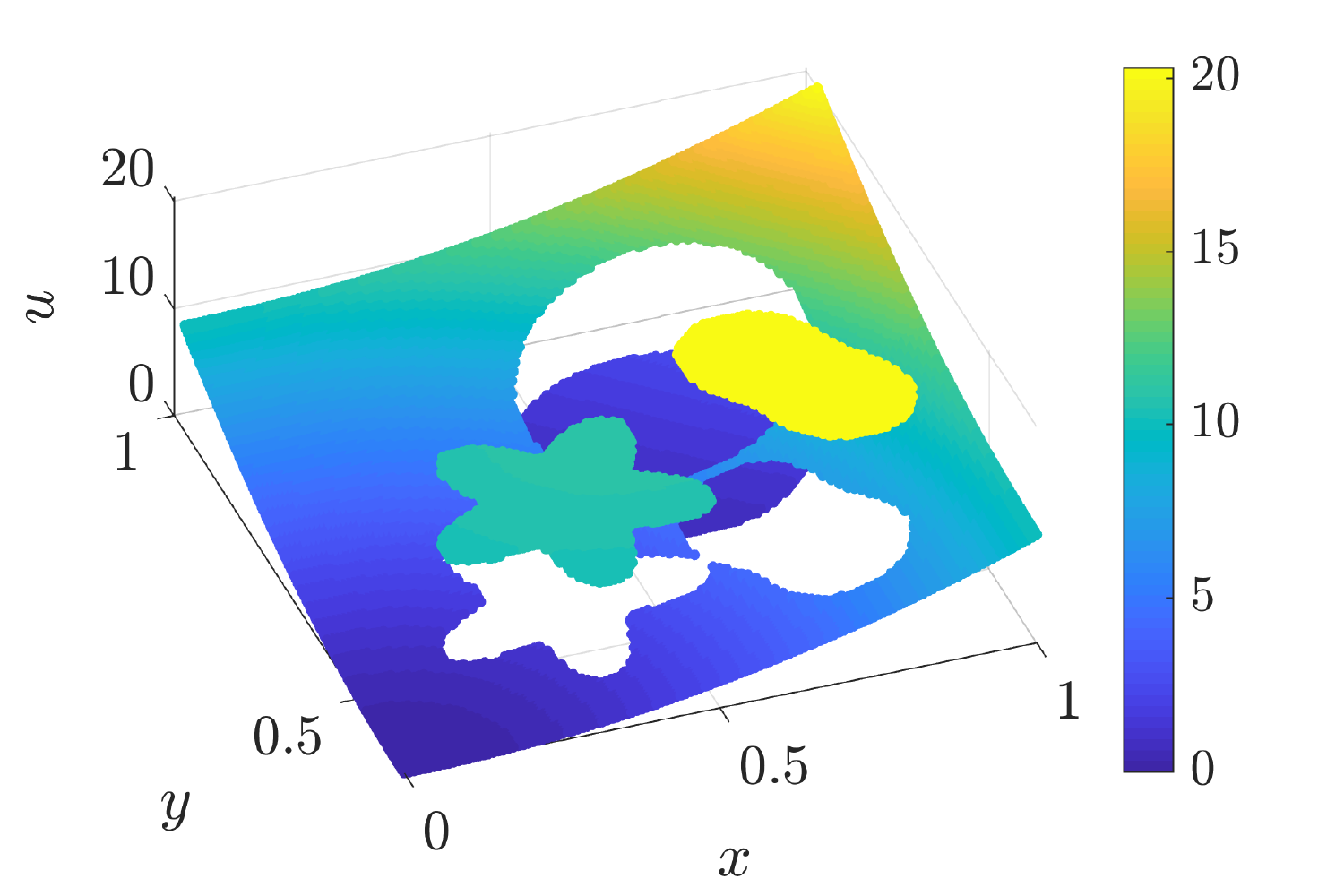}\\
 \end{center}
 \vspace{-0.2in}
 \caption{Example 3.
          Left: interfaces immersed into $G_P\/$.
          The grid $G\gn{2}\/$ used to represent $\phi_2\/$
          is also shown.
          Right: solution obtained with the CFM.}
 \label{fig:3int:grid}
\end{figure} 

\begin{figure}[t!] 
 \begin{center}
  \includegraphics[width = 3.5in]{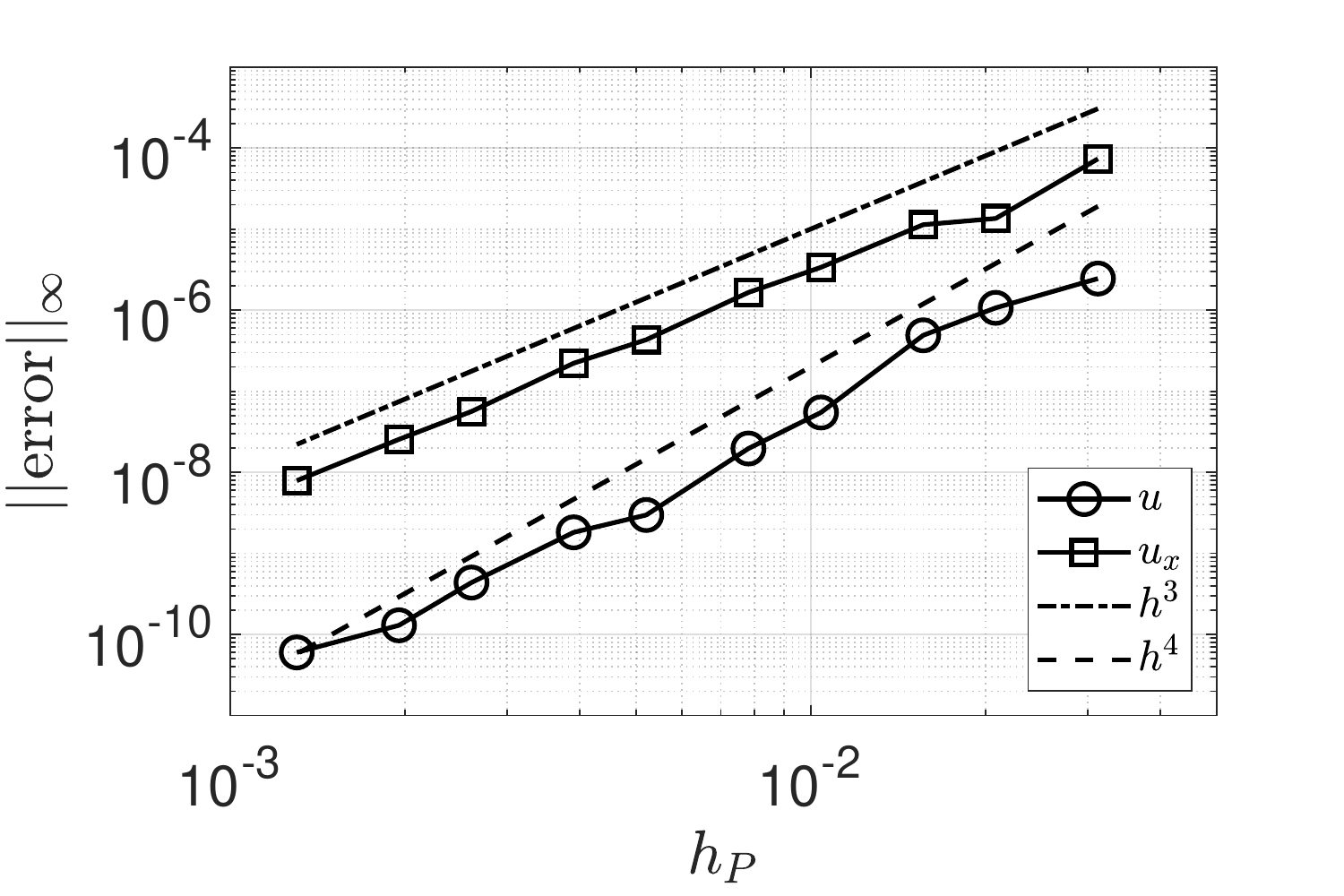}
 \end{center}
 \vspace{-0.2in}
 \caption{Example 3.
          Convergence of the error of in the solution and
          its $x$-derivative in the $L^{\infty}\/$ norm.
          The $y$-derivative behaves similarly.}
 \label{fig:3int:error}
\end{figure} 

We illustrate the concept of independent level set grids in
figure~\ref{fig:3int:grid}(left).
In this figure we show $G\gn{2}\/$ laid over $G_P\/$, along
with the immersed interfaces.
The solution obtained with the CFM is shown in
figure~\ref{fig:3int:grid}(right).
Once again we observe that the CFM produces good results in
the presence of multiple interfaces.
Furthermore, in addition to the expected fourth order
convergence of the error in the solution, we also observe
third order convergence of the error in the gradient, as
shown in figure~\ref{fig:3int:error}.

\subsection{Sphere with bumps} \label{sub:results:bumps}
\begin{figure}[t!] 
 \begin{center}
  \includegraphics[width = 3.0in]{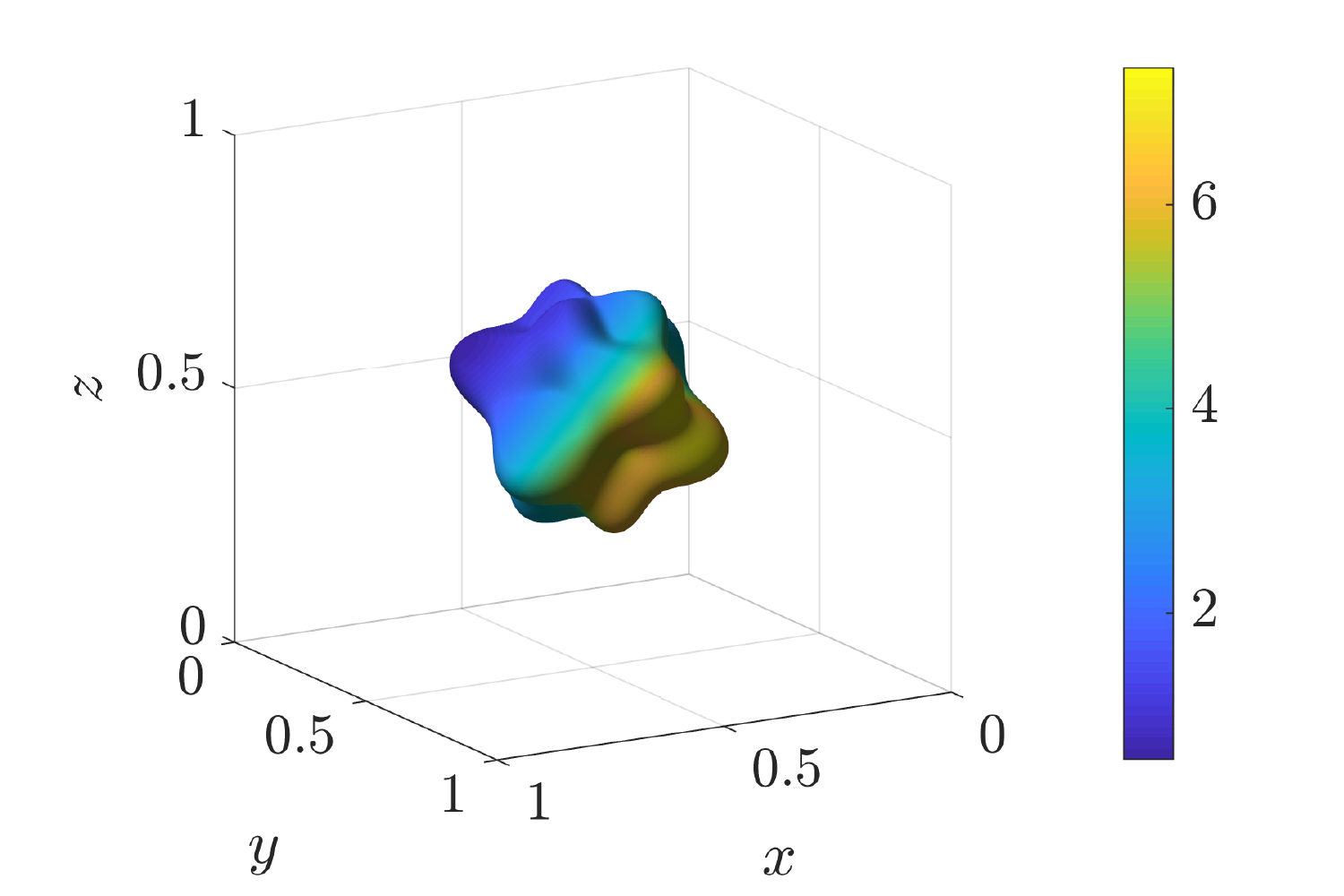}
 \end{center}
 \vspace{-0.2in}
 \caption{Example 4.
    Interface immersed into $G_P\/$, colored according to
    the intensity of the jump in the solution across it.}
 \label{fig:bumps:gamma}
\end{figure}

%
In this example we solve the following 3-D problem, which is
illustrated in figure~\ref{fig:bumps:gamma}.

\begin{itemize}
 \item $\phi(x_1\/,\,y_1\/,\,z_1) =
        (x_1-0.35)^2 + (y_1-0.35)^2 + (z_1-0.35)^2
        - r^2\/$.
 \item $r(\varphi\/,\,\psi) = \sqrt{3}/8 + 
        (0.28/\pi)(\varphi-\varphi^2/\pi)
        \sin(3\varphi)\sin(4\psi)\/$.
 \item $\varphi(x_1\/,\,y_1) =
        \arctan\Big(\dfrac{y_1-0.35}{x_1-0.35}\Big)\/$.
 \item $\psi(x_1\/,\,y_1\/,\,z_1) =
        \arccos\Bigg(\dfrac{z_1-0.35}
        {\sqrt{(x_1-0.35)^2+(y_1-0.35)^2+(z_1-0.35)^2}}
        \Bigg)\/$.
 \item $\Omega_1 = \{(x_1\/,\,y_1\/,\,z_1) \in [0,0.7]^3 \mid
        \phi(x_1\/,\,y_1\/,\,z_1) \le 0\}\/$.
 \item $\Omega_2 = \{(x\/,\,y\/,\,z) \in [0,1]^3 \mid
        \phi(x\/,\,y\/,\,z) > 0\}\/$.
 \item $u_1(x\/,\,y\/,\,z) = \sin(\pi(x+z)/\sqrt{2})\exp(\pi y)\/$.
 \item $u_2(x\/,\,y\/,\,z) = 0\/$.
\end{itemize}

In the expressions above, the level set $\phi\/$ is defined
in terms of the Cartesian coordinates aligned with $G\g\/$.
These coordinates are given by
\begin{subequations}\label{eq:bumps:x1}
 \begin{align*}
  x_1 &=  x - 0.13\/,\\
  y_1 &=  y - 0.1\/,\\
  z_1 &=  z - 0.15\/.
 \end{align*}
\end{subequations}
%
Note that $G\g\/$ is distinct from $G_P\/$, and the
relationship between grid spacings is $h\g/h_P = 0.7$.

\begin{figure}[htb!] 
 \begin{center}
  \includegraphics[height = 1.32in]{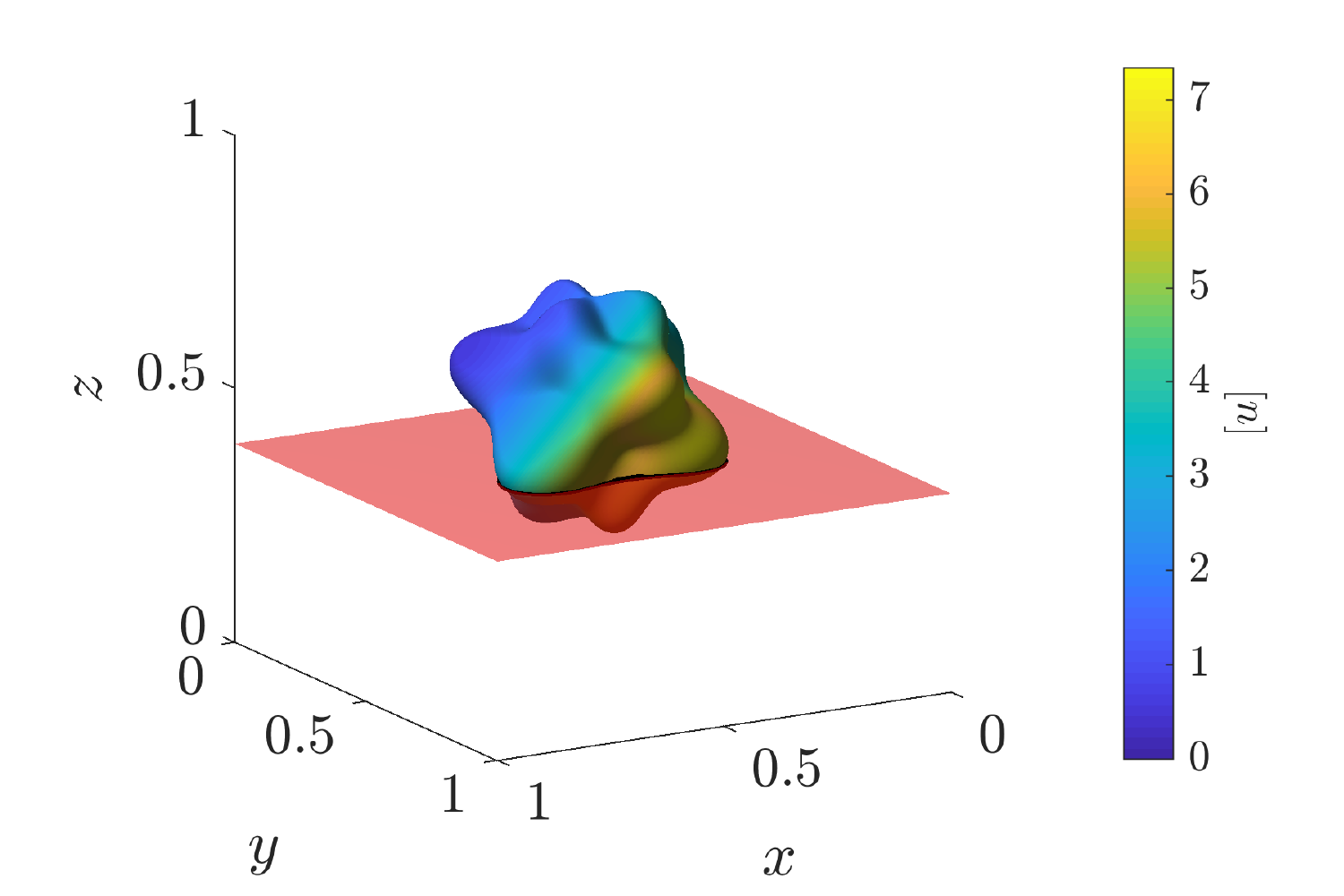}
  \includegraphics[height = 1.32in]{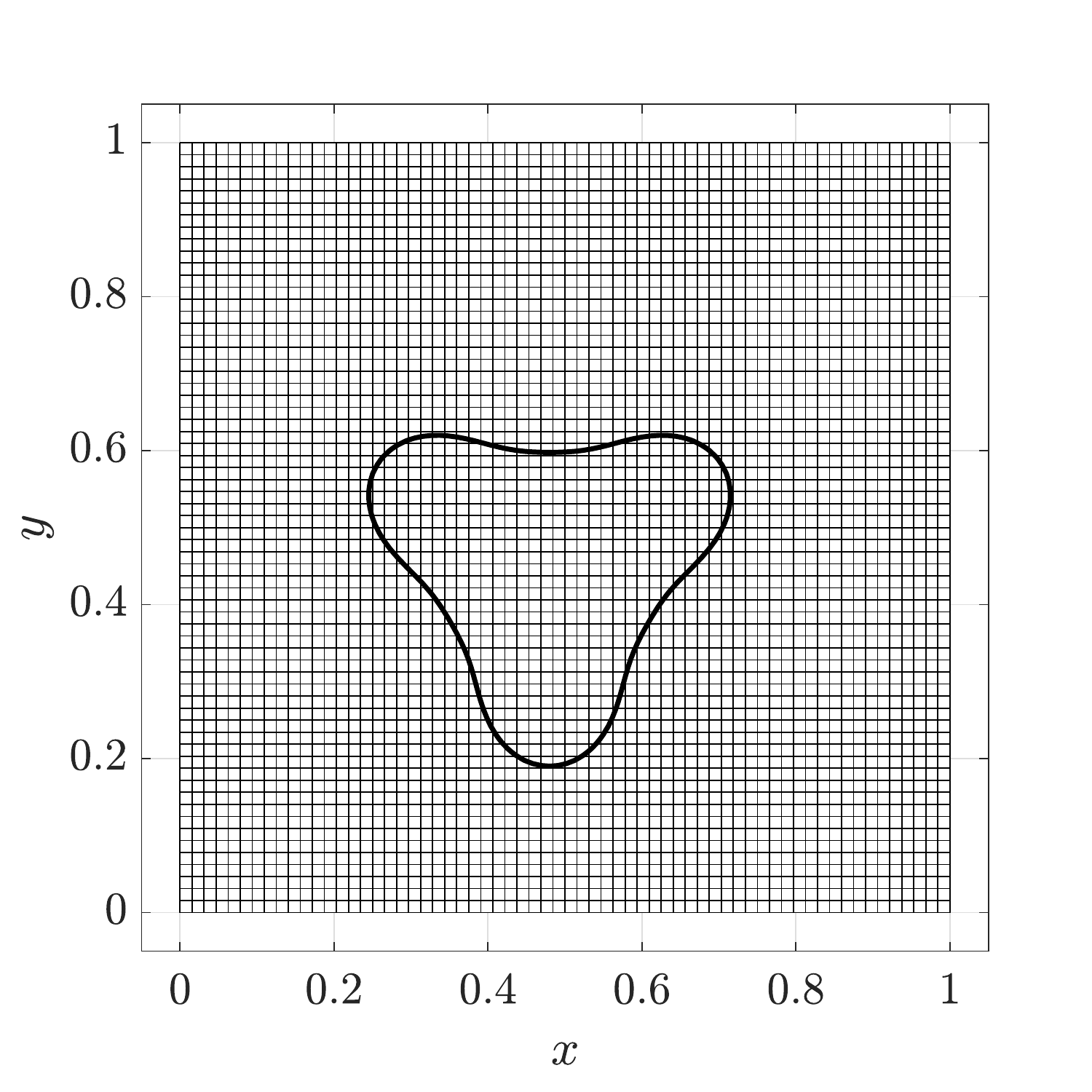}
  \includegraphics[height = 1.32in]{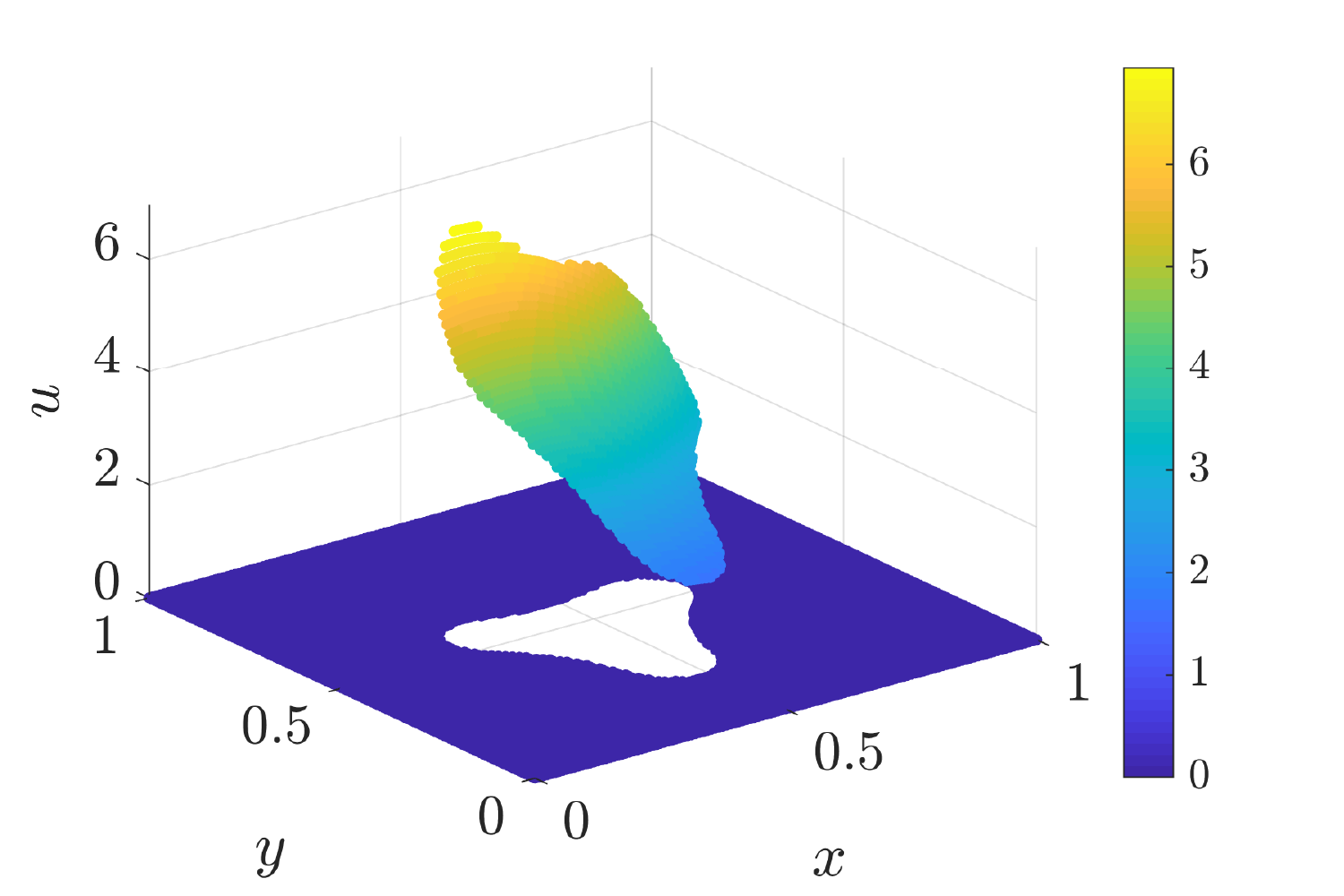}
 \end{center}
 \vspace{-0.2in}
 \caption{Example 4.
          2-D slice of the 3-D solution.
          Left: location of the slicing plane ($z=0.39$).
          Center: grid $G_P\/$.
          Right: solution obtained with the CFM.}
 \label{fig:bumps:slice}
\end{figure}

\begin{figure}[htb!] 
 \begin{center}
  \includegraphics[width = 3.5in]
   {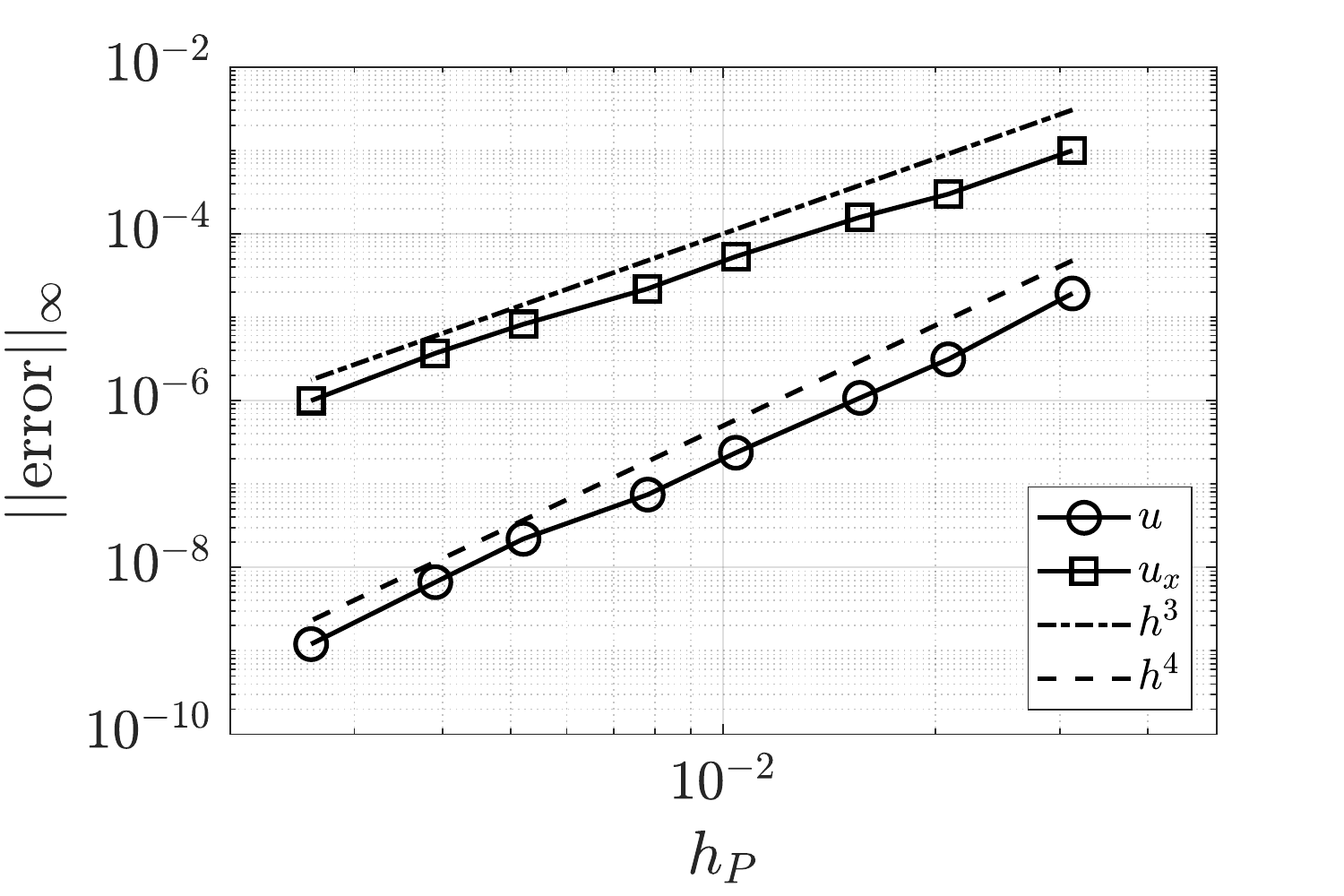}
 \end{center}
 \vspace{-0.2in}
 \caption{Example 4.
    Convergence of the error in the solution and its
    $x$-derivative in the $L^{\infty}\/$ norm.
    The $y\/$ and $z$-derivatives behave similarly.}
 \label{fig:bumps:convergence}
\end{figure}

Figure~\ref{fig:bumps:gamma} shows the interface, colored
according to the intensity in the jump in the
solution across it.
Figure~\ref{fig:bumps:slice} shows a 2-D slice of the 3-D
solution computed with the CFM, as well as the interface
immersed into the Cartesian grid at the slicing plane.
The error of the solution and its gradient are plotted in
figure~\ref{fig:bumps:convergence}.
Just as in 2-D, the accuracy is fourth order in the solution
and third order in the gradient.

\subsection{Touching spheres}
\label{sub:results:spheres}
%
Here we present a 3-D version of example 2 in
\S\ref{sub:results:circles}.
We consider two spherical interfaces that touch at a single
point, as shown in figure~\ref{fig:spheres:gamma}.
The interfaces are each represented using independent level
set grids and the relationships between grid spacings are
$h\gn{1}/h_P = 0.8$ and $h\gn{2}/h_P = 0.5$.
The problem is defined as follows:
%
\begin{figure}[t!] 
 \begin{center}
  \includegraphics[width = 3.0in]{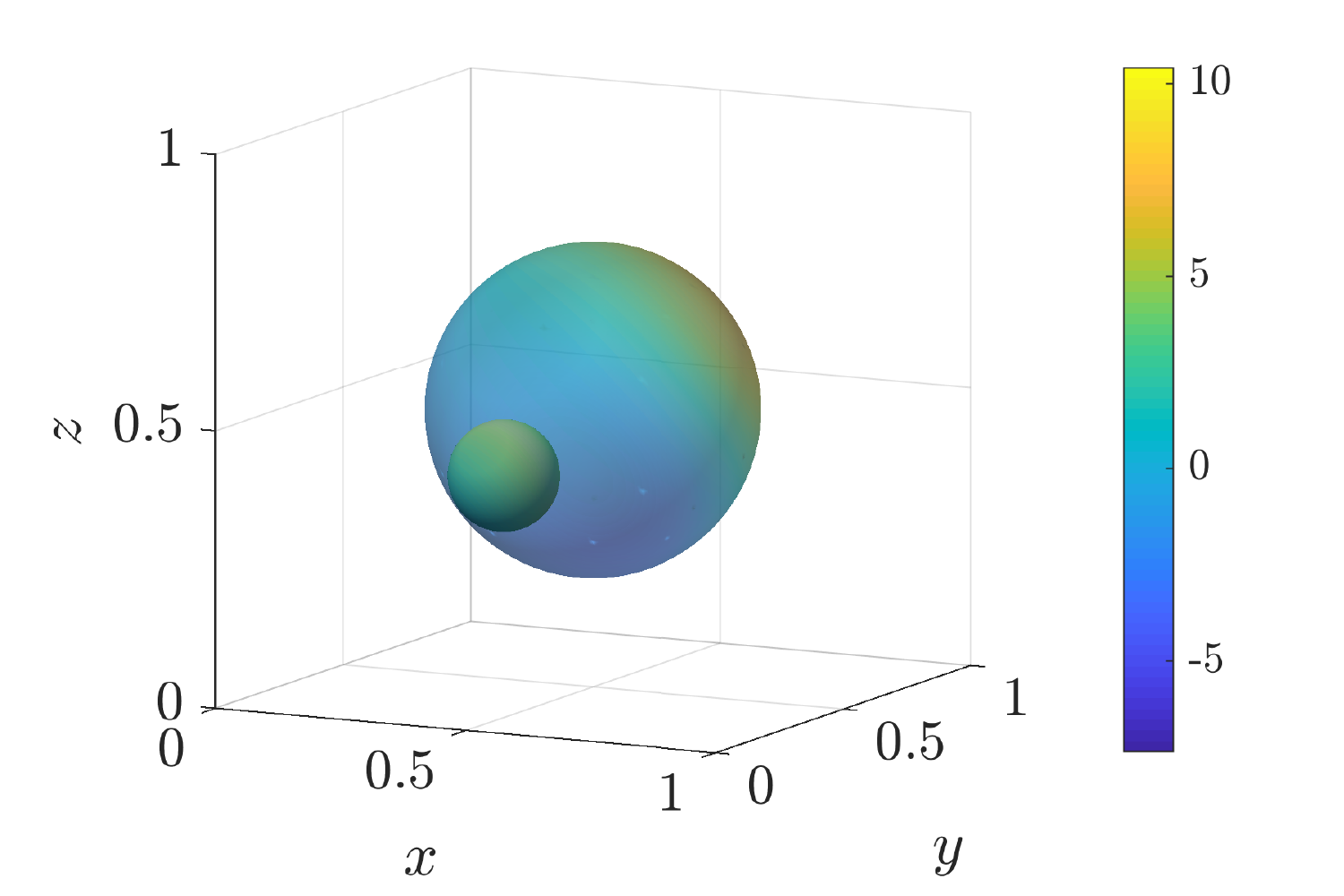}
 \end{center}
 \vspace{-0.2in}
 \caption{Example 5.
    Interfaces immersed into $G_P\/$,
    colored according to
    the intensity of the jumps in the solution across them.    
    The spheres touch at a single point.      
    Transparency is applied to the outer interface to
    show the internal interface.}
 \label{fig:spheres:gamma}
\end{figure} 
%
\begin{itemize}
 \item $\phi_1(x_1\/,\,y_1\/,\,z_1) = (x_1-0.40)^2 + (y_1-0.40)^2 +
        (z_1-0.40)^2 - 0.09\/$.
 \item $\phi_2(x_2\/,\,y_2\/,\,z_2) = (x_2-0.25)^2 + (y_2-0.25)^2 +
        (z_2-0.25)^2 - 0.01\/$.
 \item $\Omega_1 = \{(x_1\/,\,y_1\/,\,z_1) \in [0,0.8]^3 \mid,
        \phi_1(x_1\/,\,y_1\/,\,z_1) \le 0\}\/$.
 \item $\Omega_2 = \{(x_2\/,\,y_2\/,\,z_2) \in [0,0.5]^3 \mid
        \phi_2(x_2\/,\,y_2\/,\,z_2) \le 0\}\/$.
 \item $\Omega_3 = \{(x\/,\,y\/,\,z) \in [0,1]^3 \mid
        \phi_1(x\/,\,y\/,\,z) > 0, \phi_2(x\/,\,y\/,\,z) > 0\}\/$.
 \item $u_1(x\/,\,y\/,\,z) = \bigl(\sin(\pi x)\sin(\pi y) + 5
                     \bigr)\log(x + z + 2)\/$.
 \item $u_2(x\/,\,y\/,\,z) = \sin\bigl(\pi (x+z)\bigr)\bigl(\sin(\pi y)
        - \exp(\pi y)\bigr)\/$.
 \item $u_3(x\/,\,y\/,\,z) = \exp(x)\bigl(x^2\sin(y) +
        y^2\bigr)\cos(\pi z)\/$.
\end{itemize}
In the expressions above, the level sets are defined in
terms of the respective grid coordinates, which are given by
\begin{subequations}\label{eq:spheres:xi}
 \begin{align*}
  x_1 &= x - 0.1\/,\\
  y_1 &= y - 0.1\/,\\
  z_1 &= z - 0.1\/,\\
  x_2 &= x - 0.25 - 0.2\cos(\pi/e^2) \cos(\pi/3\varphi)\/,\\
  y_2 &= y - 0.25 - 0.2\sin(\pi/e^2) \cos(\pi/3\varphi)\/,\\
  z_2 &= z  - 0.25 - 0.2\sin(\pi/3\varphi)\/,
 \end{align*}
\end{subequations}
where $\varphi\/$ denotes the golden ratio,
$\varphi = \frac{1 + \sqrt{5}}{2}\/$.

\begin{figure}[htb!] 
 \begin{center}
  \includegraphics[height = 1.32in]
   {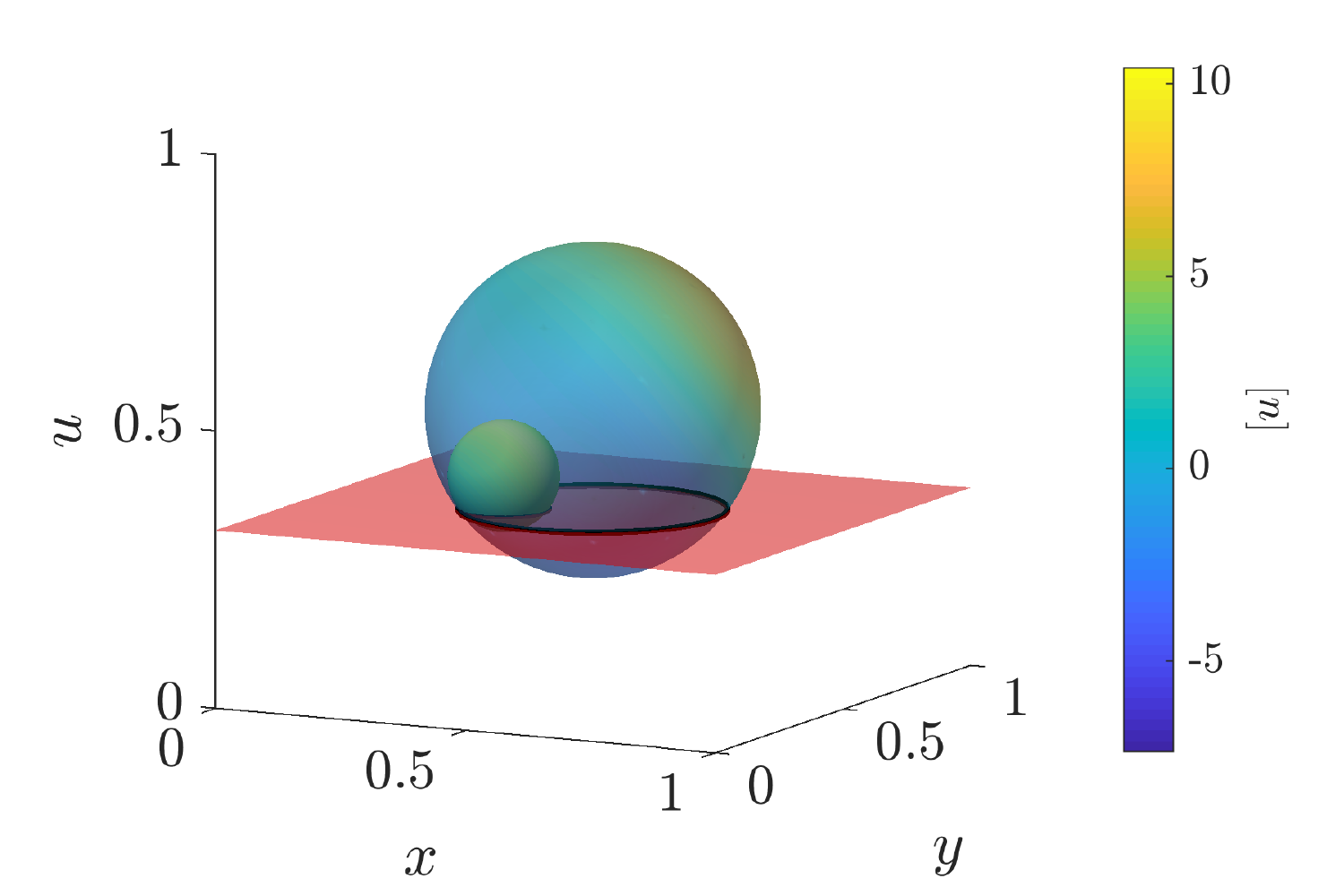}
  \includegraphics[height = 1.32in]
   {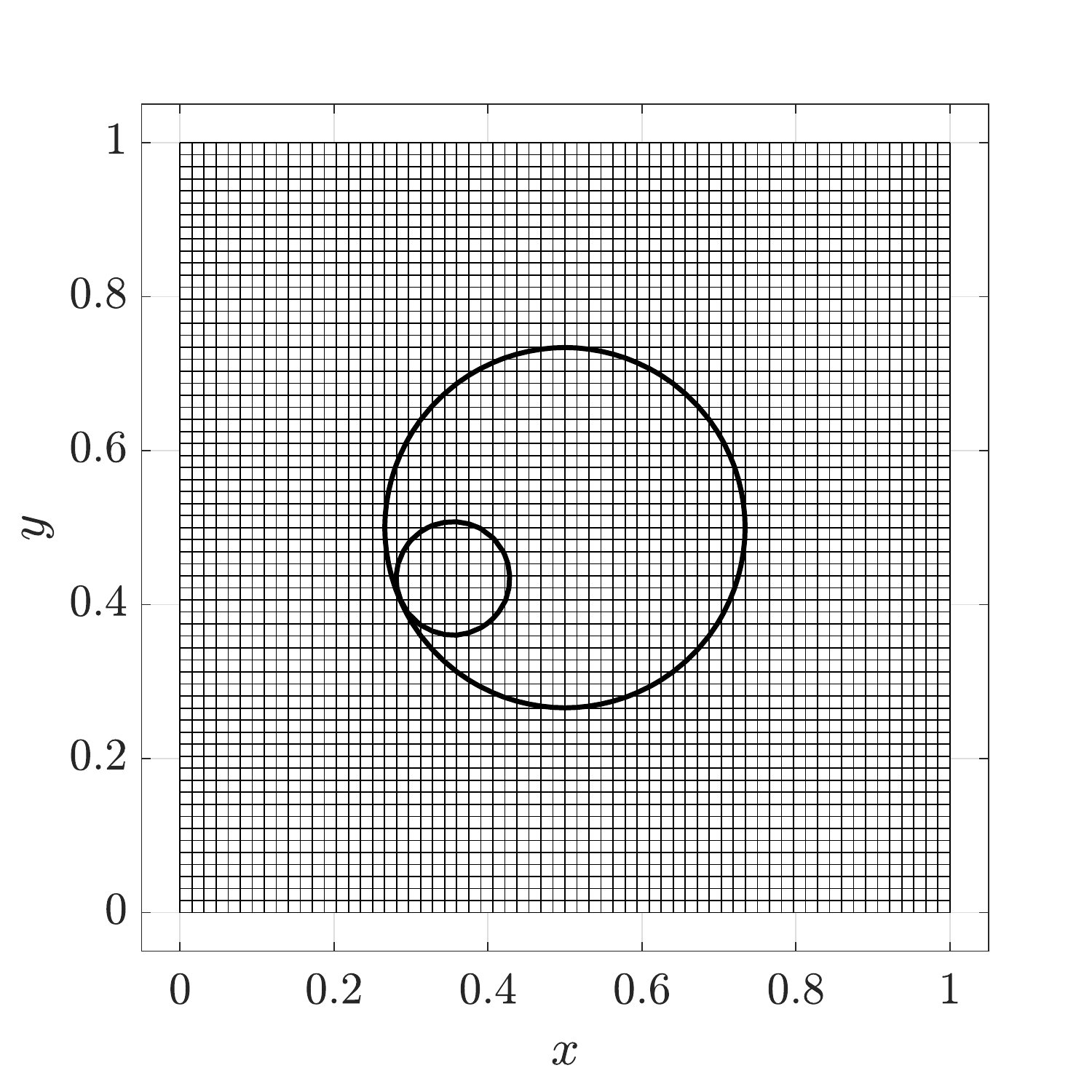}
  \includegraphics[height = 1.32in]
   {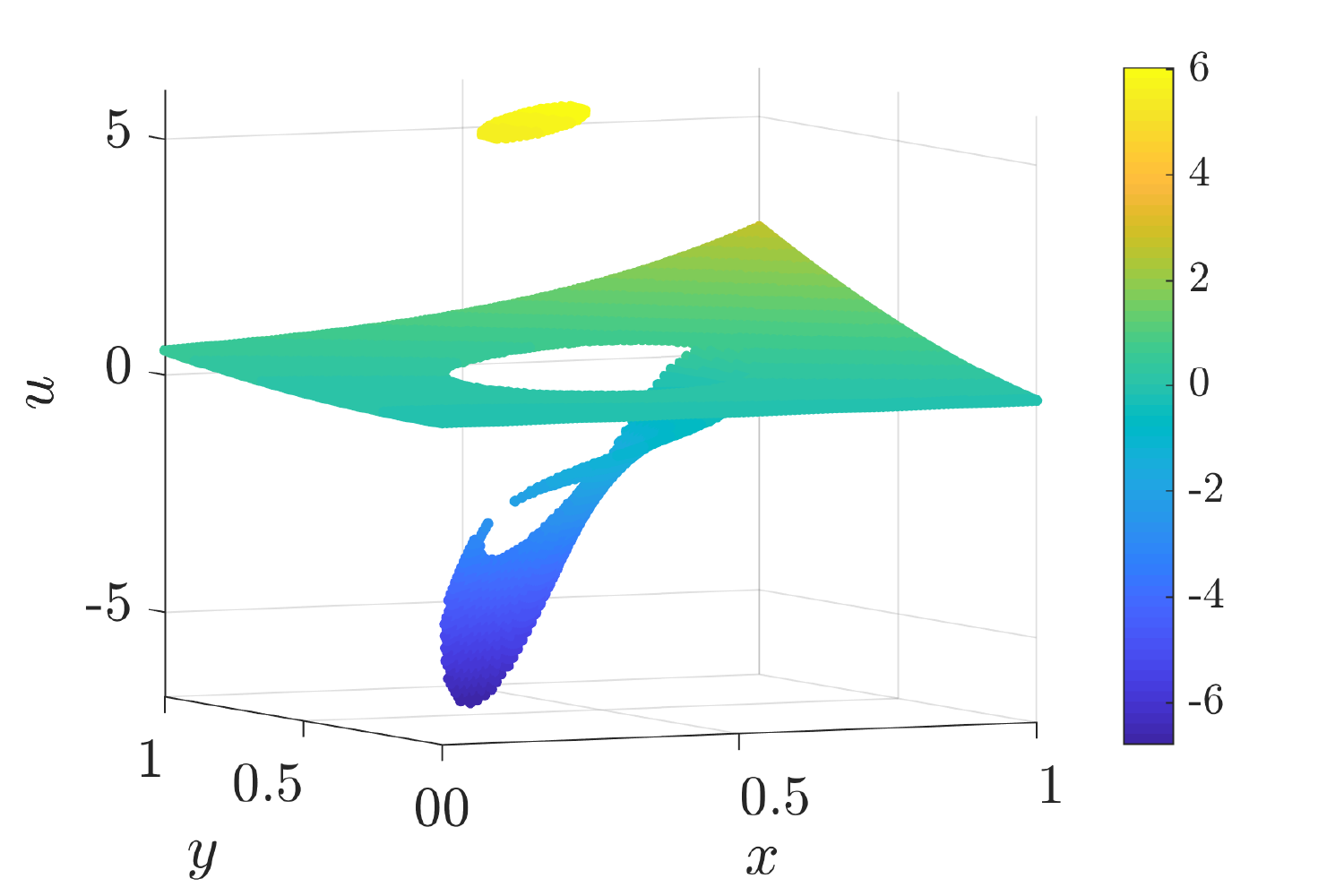}
 \end{center}
 \vspace{-0.2in}
 \caption{Example 5.
          2-D slice of the 3-D solution.
          Left: location of the slicing plane ($z=0.32$).
          Center: grid $G_P\/$.
          Right: solution obtained with the CFM.}
 \label{fig:spheres:slice}
\end{figure}

\begin{figure}[htb!] 
 \begin{center}
  \includegraphics[width = 3.5in]{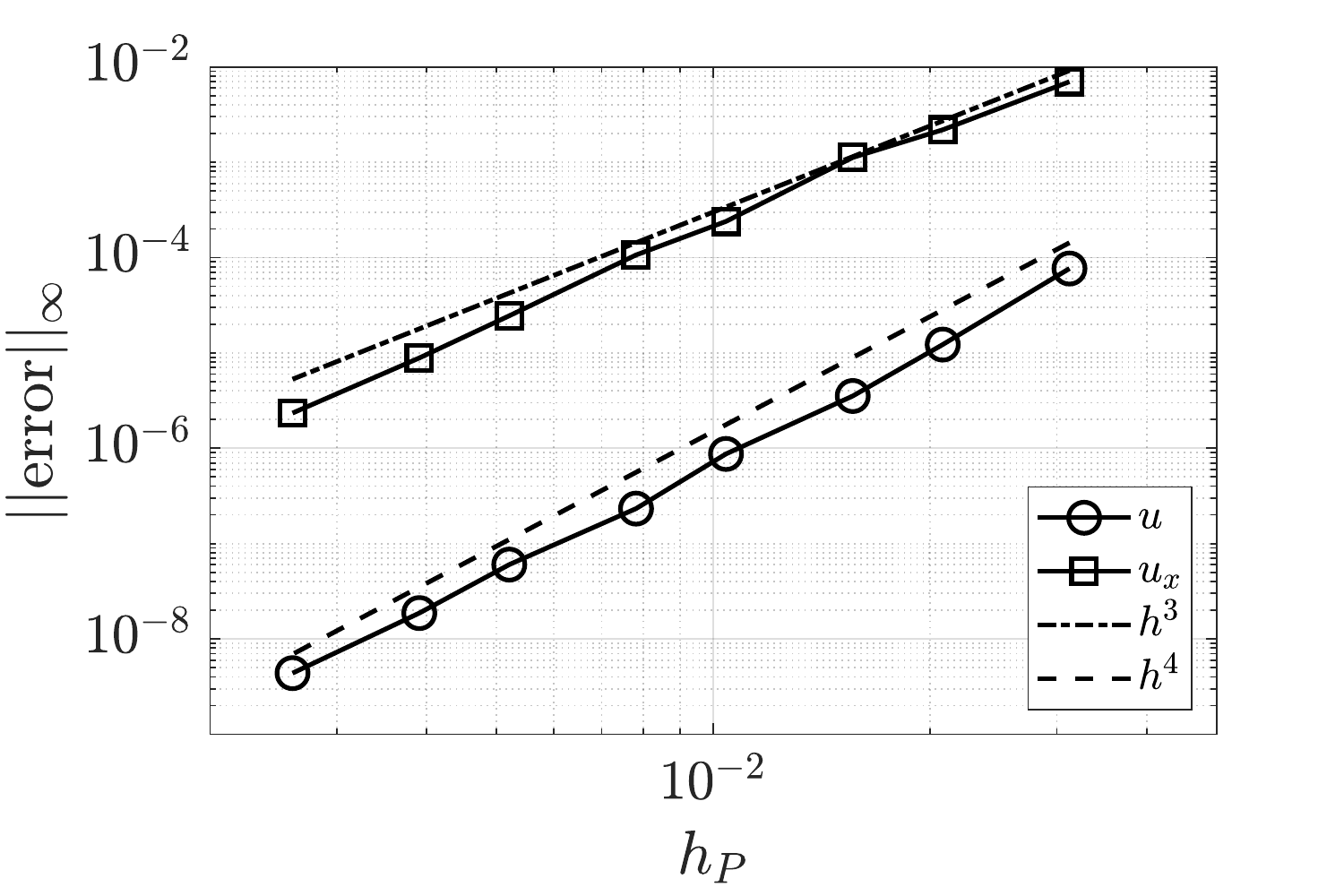}
 \end{center}
 \vspace{-0.2in}
 \caption{Example 5.
    Convergence of the error in the solution and its
    $x$-derivative in the $L^{\infty}\/$ norm. The $y\/$ and
    $z$-derivatives behave similarly.}
 \label{fig:spheres:convergence}
\end{figure}

Figure~\ref{fig:spheres:gamma} shows the interfaces, 
colored according to the intensity of the jumps in the
solution across them.
Figure~\ref{fig:spheres:slice} shows a 2-D slice of the 3-D
solution, as computed with the CFM, on a plane close to the
contact point between the spheres.
The error in the solution and gradient are plotted in
figure~\ref{fig:spheres:convergence}.
Note that the accuracy of the solution and gradient
\textit{does not degrade}, even though the interfaces are
arbitrarily close.

\subsection{3-D interfaces touching at two points}
\label{sub:results:2int}
In this last example we apply the CFM to a 3-D problem with
two interfaces that touch at two points, as shown in
figure~\ref{fig:2int:gamma}.
The interfaces are represented using separate level set
grids and the relationships between the grid spacings are
$h\gn{1}/h_P = 0.85$ and $h\gn{2}/h_P = 0.6$.
The problem is defined as follows:
%
\begin{figure}[htb!] 
 \begin{center}
  \includegraphics[width = 3.0in]{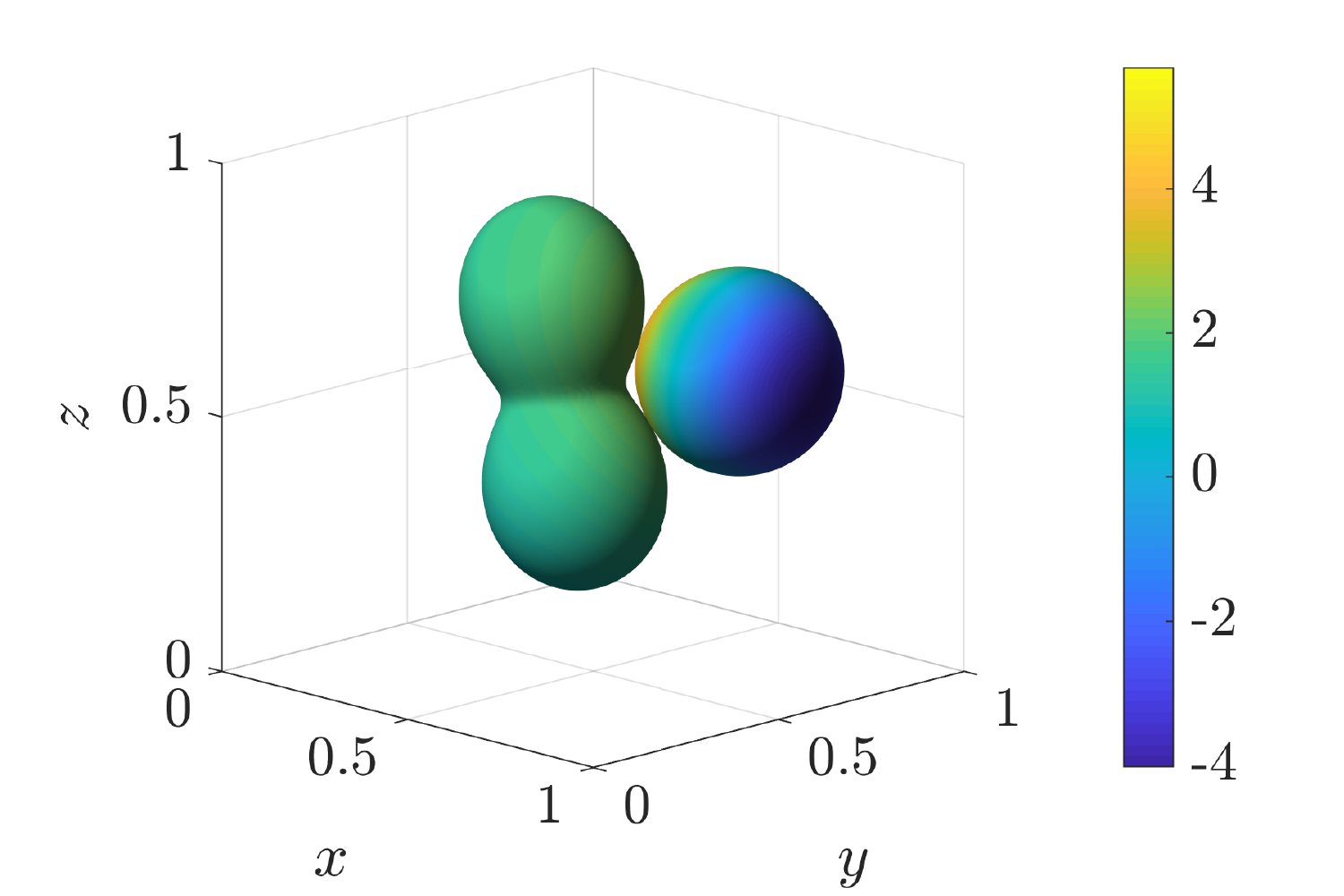}
 \end{center}
 \vspace{-0.2in}
 \caption{Example 6.
    Interfaces immersed into $G_P\/$, colored according to
    the intensity of the jumps in the solution across them.
    The interfaces touch at two distinct points.}
 \label{fig:2int:gamma}
\end{figure} 
%
\begin{itemize}
 \item $R = \sqrt{(x_1-0.425)^2 + (y_1-0.425)^2 + (z_1-0.425)^2}\/$.
 \item $\psi(x_1\/,\,y_1\/,\,z_1) =
        \arccos\Biggl(\dfrac{z_1-0.425}{R^2}\Biggr)\/$.
 \item $r(\psi) = 0.25 + 0.13\cos(2\psi)\/$.
 \item  $\phi_1(x_1\/,\,y_1\/,\,z_1) = R^2-r^2\/$
 \item $\phi_2(x_2\/,\,y_2\/,\,z_2) = (x_2-0.3)^2 + (y_2-0.3)^2 +
        (z_2-0.3)^2 - 0.04\/$.
 \item $\Omega_1 = \{(x_1\/,\,y_1\/,\,z_1) \in [0,0.85]^3 \mid
        \phi_1(x_1\/,\,y_1\/,\,z_1) \le 0\}\/$.
 \item $\Omega_2 = \{(x_2\/,\,y_2\/,\,z_2) \in [0,0.6]^3 \mid
        \phi_2(x_2\/,\,y_2\/,\,z_2) \le 0\}\/$.
 \item $\Omega_3 = \{(x\/,\,y\/,\,z) \in [0,1]^3 \mid
        \phi_1(x\/,\,y\/,\,z) > 0, \phi_2(x\/,\,y\/,\,z) > 0\}\/$.
 \item $u_1(x\/,\,y\/,\,z) = x\,y\,z\/$.
 \item $u_2(x\/,\,y\/,\,z) = \exp(x)\cos(y)\sin(z)\/$.
 \item $u_3(x\/,\,y\/,\,z) = \log\bigl(\sqrt{(x-0.661)^2
        + (y-0.651)^2 + (z-0.636)^2}\bigr)\/$.
\end{itemize}
Note that, in the expressions above, the level set functions
are defined in terms of their respective grid coordinates,
given by
\begin{subequations}\label{eq:2int:xi}
 \begin{align*}
  x_1 &= (x - 0.1)\cos(10\pi /180)
       + (z - 0.05)\sin(10\pi /180)\/,\\
  y_1 &= y - 0.05\/,\\
  z_1 &= -(x - 0.1)\sin(10\pi /180)
       + (z - 0.05)\cos(10\pi /180)\/,\\
  x_2 &= x - 0.1047\/,\\
  y_2 &= y - 0.1842\/,\\
  z_2 &= z - 0.1249\/.
 \end{align*}
\end{subequations}

\begin{figure}[htb!] 
 \begin{center}
  \includegraphics[height = 1.32in]
   {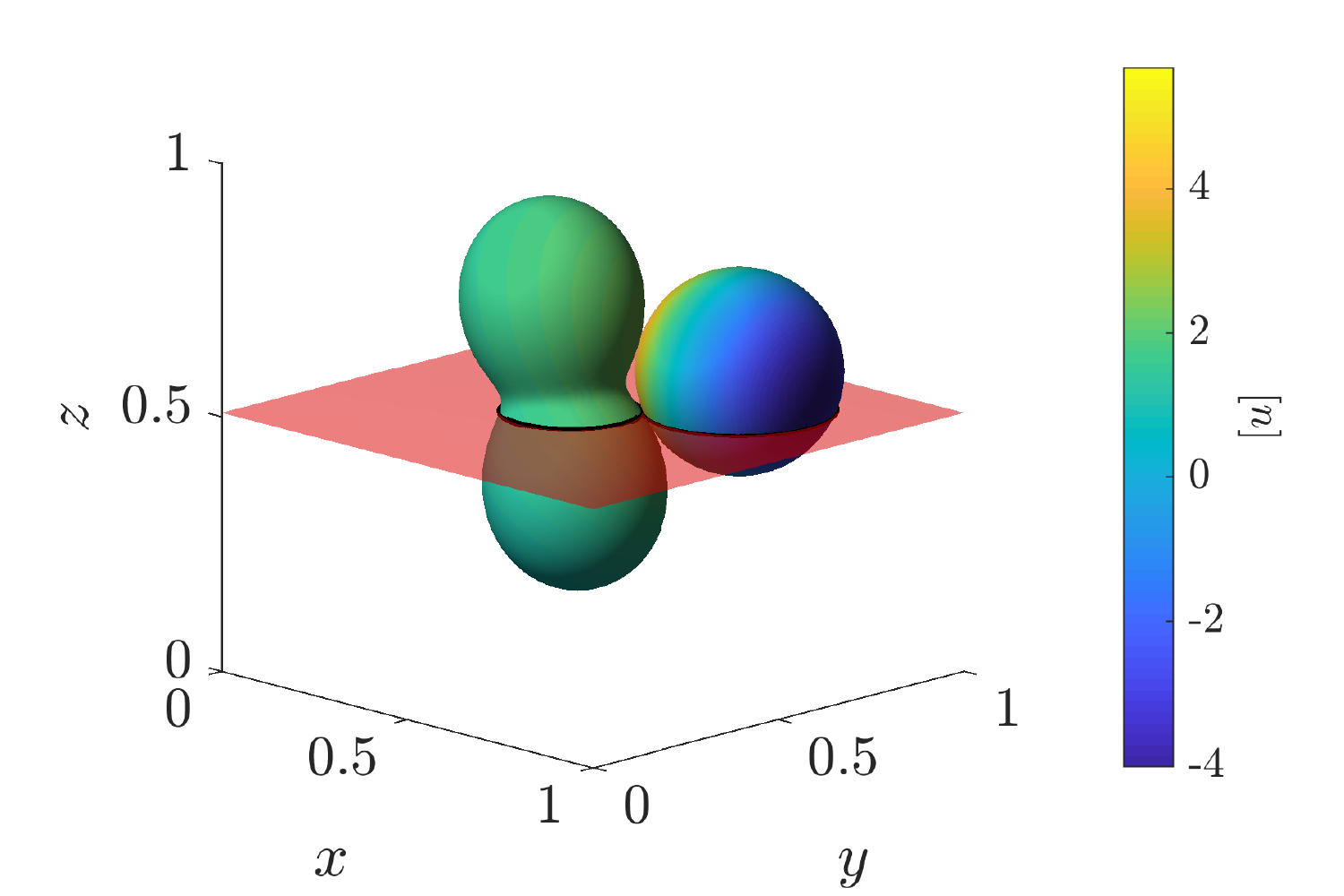}
  \includegraphics[height = 1.321in]
   {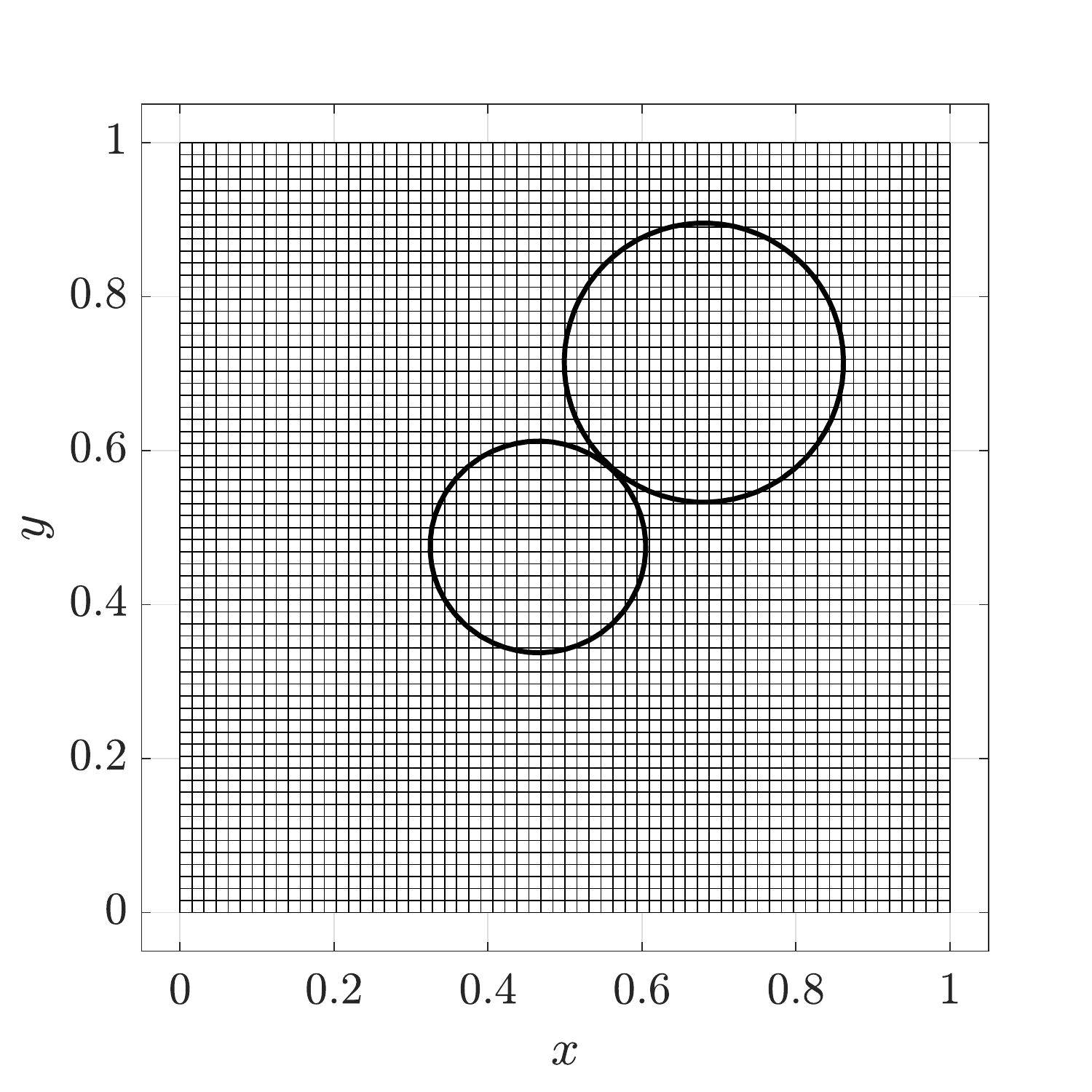}
  \includegraphics[height = 1.32in]
   {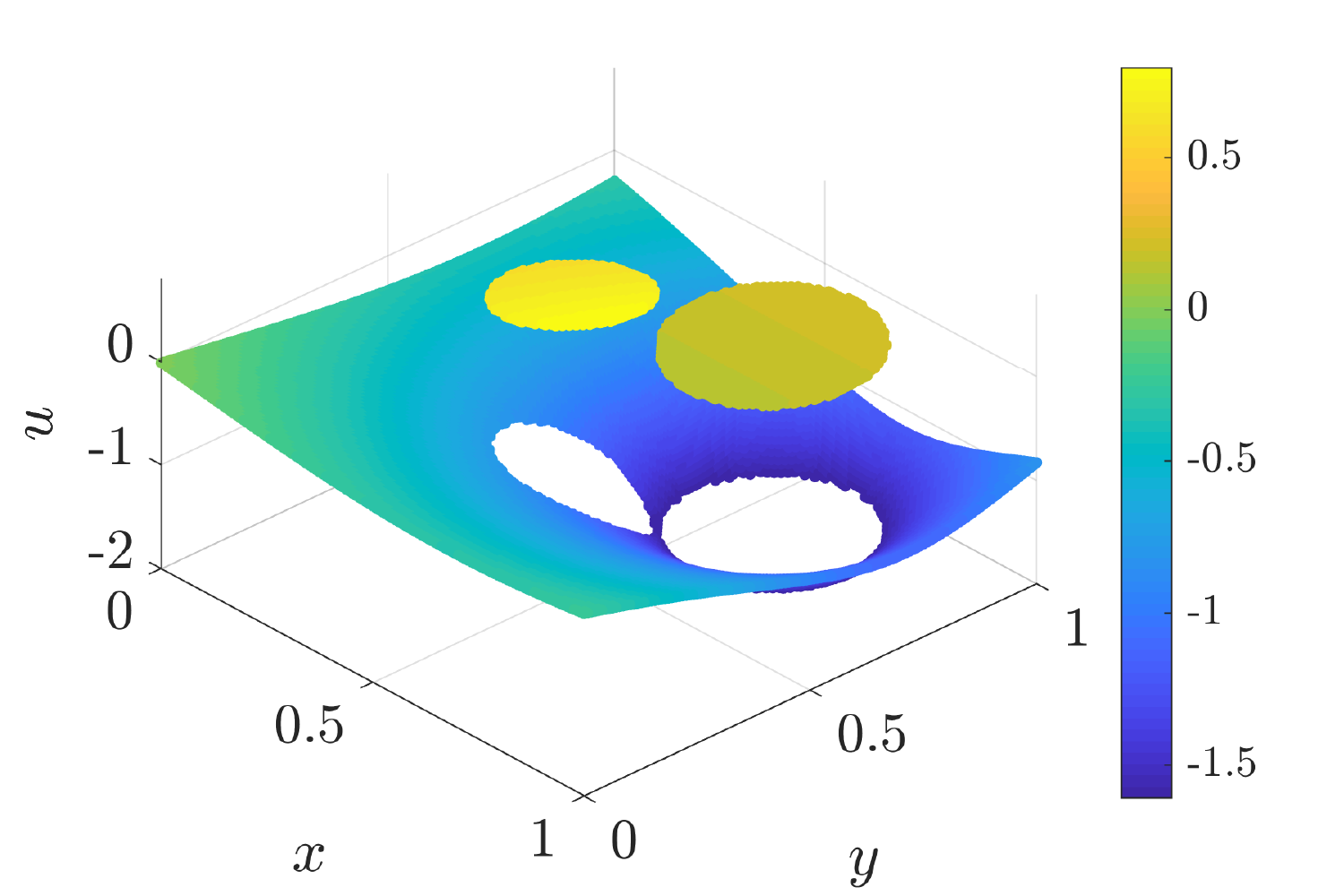}
 \end{center}
 \vspace{-0.2in}
 \caption{Example 6.
          2-D slice of the 3-D solution.
          Left: location of the slicing plane ($z=0.5$).
          Center: grid $G_P\/$.
          Right: solution obtained with the CFM.}
 \label{fig:2int:slice}
\end{figure}

\begin{figure}[htb!] 
 \begin{center}
  \includegraphics[width = 3.5in]{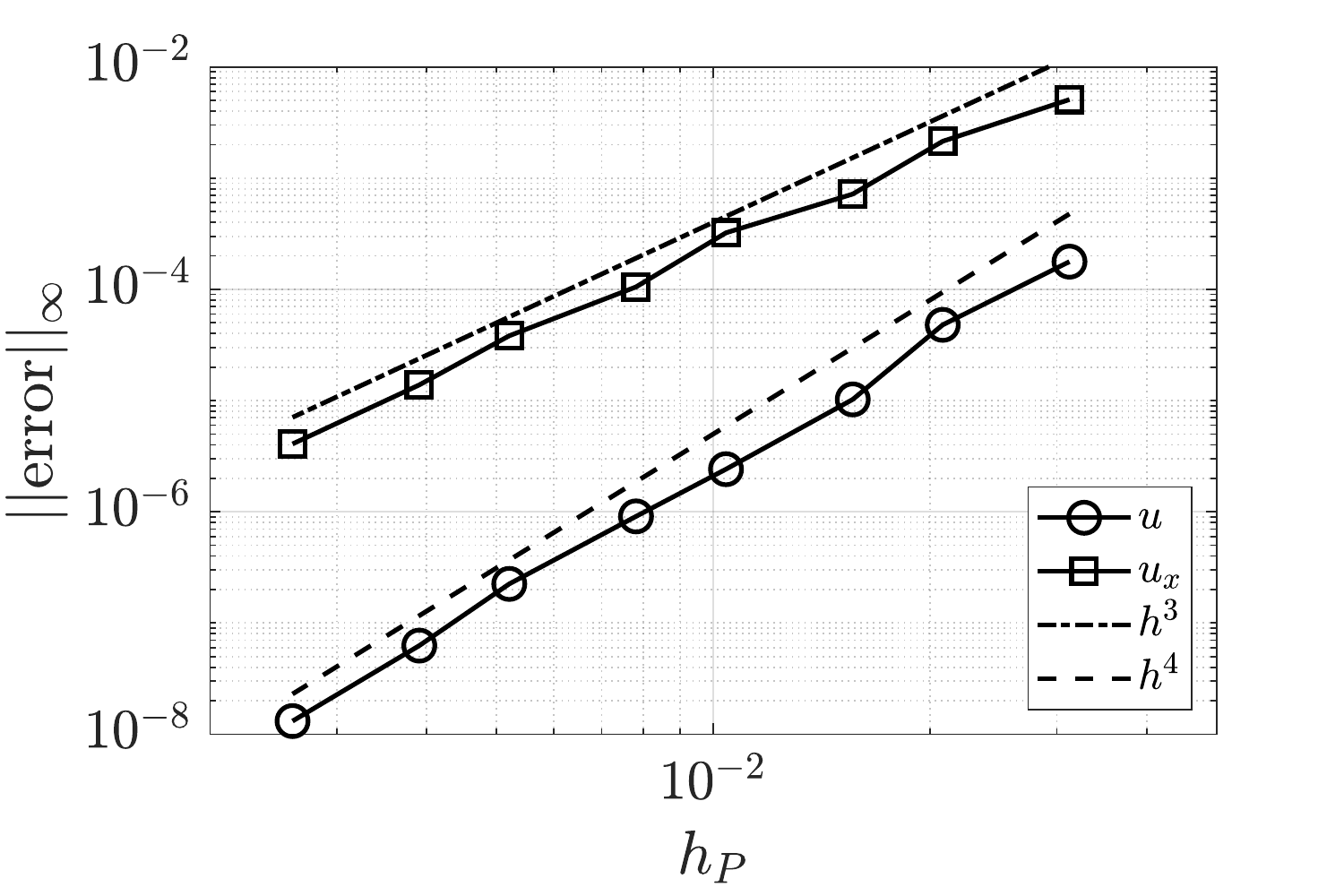}
 \end{center}
 \vspace{-0.2in}
 \caption{Example 6.
    Convergence of the error in the solution and its
    $x$-derivative in the $L^{\infty}\/$ norm. The other
    derivatives behave similarly.}
 \label{fig:2int:convergence}
\end{figure}

Figure~\ref{fig:2int:gamma} shows the interfaces, colored
according to the intensity of the jump in the solution
across them.
Figure~\ref{fig:2int:slice} shows a 2-D slice of the 3-D
solution, as computed with the CFM, on a plane close to one
of the contact points.
The errors in the solution and its gradient are plotted in
figure~\ref{fig:2int:convergence}.
These results corroborate the accuracy of this CFM
implementation, as well as its robustness with respect to
situations with arbitrarily close interfaces.
%
%
%


\section{Conclusion} \label{sec:conclusion}
We present a new technique to impose jump conditions in a
least squares sense in the context of the
\textit{Correction Function Method} (CFM).
This technique results in a re-formulation of the CFM that
significantly simplifies numerical evaluation of surface
integrals associated with the jump conditions, especially
when only an implicit representation of the interface is
available (e.g., the zero contour of a level set function).
Furthermore, this new formulation preserves the main
features of the CFM: high order of accuracy and compact
discretization stencils.

The technique introduced here uses approximate coordinate
transformations to map $L^{\infty}\/$ balls on the interface
onto squares, defining interface sections over which
integration can be carried out easily using standard
numerical quadrature.
The technique also exploits the flexibility of the CFM
framework to create a least squares formulation that only
involves integrals over collections of such $L^{\infty}\/$
balls, resulting in simple integral evaluations.

We show with numerical experiments that the reformulation of
the CFM incorporating the new technique is efficient,
accurate, and robust with respect to the arbitrary fashion
in which the interface can intersect the computational grid.
In particular, we show fourth order of accuracy when solving
Poisson's equation~\eqref{eq:poisson} when the diffusion
coefficient $\beta\/$ is constant.
An extension of this approach to problems involving
discontinuous $\beta\/$ is the subject of current work.
%



\section*{Acknowledgements}

The authors are thankful to comments offered by the
anonymous reviewers.
The research of A.~N.~Marques was partially supported by
DARPA and AFOSR-COE.
The research of J.-C.~Nave was partially
supported by the NSERC Discovery Program.
The research of R.~R.~Rosales was partially supported
by the National Science Foundation grants
DMS-1719637 and DMS-1614043.



\section*{References}

\bibliographystyle{elsarticle-num}
\bibliography{biblio}







\end{document}